\pdfoutput=1
\documentclass{aa}

\usepackage{graphicx}
\usepackage{txfonts}
\usepackage{booktabs}
\usepackage{array}
\usepackage{amssymb}
\let\linenumbers\nolinenumbers

\newcommand{\CU}{\,\mathrm{C.U.}}
\newcommand{\sens}[1]{#1\,\%\CU/\sqrt{T[\mathrm{h}]}}
\newcommand{\GbmN}{4401}
\newcommand{\GbmFirst}{14 July 2008}
\newcommand{\GbmLast}{21 October 2026}

\newcommand{\GbmYears}{18}
\newcommand{\GbmPerYear}{241}
\newcommand{\GbmRetrieved}{23 October 2026}

\newcommand{\GbmTenMinLaPalma}{0.2}
\newcommand{\GbmTenMinCTAO}{0.4}
\newcommand{\GbmTenMinArray}{1.2--1.4}
\newcommand{\GbmTenMinOneLaPalma}{0.3}
\newcommand{\GbmTenMinOneCTAO}{0.5}
\newcommand{\GbmTenMinOneArray}{1.3}
\newcommand{\GbmLaPalma}{9.0}

\newcommand{\GbmLaPalmaRate}{22}
\newcommand{\GbmCTAO}{16.6}
\newcommand{\GbmCTAOInt}{17}
\newcommand{\GbmCTAORate}{40}
\newcommand{\GbmTen}{56.9}
\newcommand{\GbmTwelve}{58.8}
\newcommand{\GbmFourteen}{61.2}
\newcommand{\GbmOpt}{62.3}

\newcommand{\GbmFourteenIntOne}{61}
\newcommand{\GbmFourteenRate}{150}
\newcommand{\GbmGainDark}{3.7}

\newcommand{\GbmNotCTAO}{73}
\newcommand{\GbmCTAOMoon}{9.4}
\newcommand{\GbmFourteenMoon}{38.2--39.3}
\newcommand{\GbmGainMoon}{4.1--4.2}
\newcommand{\GbmGainZ}{4.1--4.2}
\newcommand{\GbmGainWeather}{4.3}

\newcommand{\GbmGainMoonFourteen}{4.1}
\newcommand{\GbmGainWeatherFourteen}{4.3}
\newcommand{\GbmMult}{2.1}
\newcommand{\GbmGeTwo}{40}
\newcommand{\GbmCTAOMult}{1.10}
\newcommand{\GbmLooLoss}{3.1}
\newcommand{\GbmBuildOne}{11.0}
\newcommand{\GbmBuildTwo}{20.6}
\newcommand{\GbmBuildSmallFrom}{ninth}
\newcommand{\GbmBuildSix}{47}
\newcommand{\GbmSaaDeficit}{14}
\newcommand{\GbmPacificExcess}{18}
\newcommand{\GbmSingleMin}{7.9}
\newcommand{\GbmSingleMax}{11.0}
\newcommand{\GbmIsoLess}{5}
\newcommand{\GbmIsoGainDark}{3.5}
\newcommand{\GbmIsoGainMoon}{3.9}
\newcommand{\GbmCTAOSixtyFive}{18.9}
\newcommand{\GbmFourteenSixtyFive}{66.4}
\newcommand{\GbmGainDarkSixtyFive}{3.5}
\newcommand{\GbmGainFifty}{4.1}
\newcommand{\GbmGainThirty}{5.6}
\newcommand{\MoonSepCovFourteen}{1.1}
\newcommand{\MoonSepCovCTAO}{0.2}
\newcommand{\GbmMoonSepFourteen}{1.1}
\newcommand{\GbmMoonSepCTAO}{0.2}
\newcommand{\GbmGainMoonSepFourteen}{4.0}
\newcommand{\GbmOptDiff}{0.3--1.1}
\newcommand{\GbmOptExist}{61.5}
\newcommand{\GbmOptNone}{63.3}
\newcommand{\GbmChecks}{}

\begin{document}

\title{RevolIACT: a longitudinally distributed array of compact AI-driven Cherenkov
       telescopes for the time-domain sub-TeV sky}
\titlerunning{RevolIACT: compact Cherenkov telescopes for the time-domain sub-TeV sky}
\authorrunning{D. Paneque et al.}

\author{D. Paneque\inst{1}\corrauth{dpaneque@mpp.mpg.de} \and E. Jobst\inst{1,2}
        \and L. Heinrich\inst{2} \and A. Hahn\inst{1}}
\institute{Max-Planck-Institut f\"ur Physik (Werner-Heisenberg-Institut),
           Boltzmannstr.\ 8, 85748 Garching, Germany
      \and Technical University of Munich, School of Natural Sciences,
           85748 Garching, Germany}

\date{Received ---; accepted ---}

\abstract
{The coming decade will be the golden era of time-domain astronomy. In ten years,
 however, sensitive very-high-energy observations below $\sim$1\,TeV may be
 possible only at the two sites of the Cherenkov Telescope Array Observatory
 (CTAO). Each can observe a source for only a few hours per night.}
{We quantify the loss caused by this limited coverage and assess whether RevolIACT
 (REal-sky Video-vision-transformer-Optimized Longitudinally distributed Imaging
 Atmospheric Cherenkov Telescopes), an array of compact stand-alone AI-driven
 stations, can close the gap.}
{We estimate the sensitivity that a stand-alone station needs to detect important transients seen by current imaging atmospheric Cherenkov
 telescopes (IACTs). We also compute the temporal coverage of layouts of ten
 to fourteen stations distributed in longitude, and how many of the \GbmN{}
 \textit{Fermi}-GBM gamma-ray bursts each layout could have observed when they occurred.}
{For each station, a sensitivity of $\sim$15\,\% of the Crab flux in one hour above
 0.2--0.3\,TeV, about three times worse than that of current IACT systems, would suffice for an array to detect most of these transients. With the
 standard analysis, a telescope with a 5\,m diameter mirror misses this target by a factor of 17. With a video
 vision transformer, it comes within a factor of two in an idealised simulation, and a
 6\,m telescope within 1.5. Ten to fourteen stations would provide
 three to four times the temporal coverage of CTAO, averaged over the sky and
 for the GBM bursts. Fourteen stations can observe most sources around the clock for
 about two months per year.}
{Temporal coverage, not only flux sensitivity, will limit sub-TeV transient astronomy
 in the next decade. RevolIACT would complement CTAO. The concept relies on the transformer analysis,
 which is still young and will improve. The decisive next step is to confirm this performance on real data, with a prototype station.}


\maketitle

\section{Introduction}
\label{sec:intro}

For almost four decades, very-high-energy (VHE) gamma-ray astronomy has progressed by
building larger telescopes and by observing in stereoscopic mode, in which several
telescopes observe the same air shower. The dish diameter went from the 10\,m of
Whipple \citep{kildea2007} to the 17\,m of MAGIC \citep{aleksic2016}, the 28\,m of
the largest H.E.S.S.\ telescope \citep{hess2017ct5} and the 23\,m of the
Large-Sized Telescope (LST) sub-array of the Cherenkov Telescope Array Observatory
\citep[CTAO;][]{acharya2013,abe2023}. This strategy has been extraordinarily
successful. The number of known VHE sources has increased from a handful in the
1990s to more than two hundred in the TeVCat catalogue \citep{wakely2008}. Imaging
atmospheric Cherenkov telescopes (IACTs) have detected pulsed gamma rays from the
Crab, Vela and Geminga pulsars, at energies from below 20\,GeV up to 20\,TeV
\citep{magic2008crab,veritas2011crab,magic2016crab,hess2018vela,hess2023vela,magic2020geminga,lst2025geminga}.
They have also detected flat-spectrum radio
quasars at redshifts close to 1, such as PKS\,1441+25, QSO\,B0218+357 and OP\,313
\citep{magic2015pks1441,magic2016b0218,abe2026op313}, whose gamma rays left them
when the Universe was less than half its present age. The absorption of these
gamma rays by the extragalactic background light (EBL), the light emitted by all
stars and galaxies over cosmic history, has allowed IACTs to constrain the density
of this light \citep{aharonian2006ebl,magic2008_3c279,acciari2019ebl}. IACTs have also
measured flux variations of active galactic nuclei (AGN) on timescales of minutes
\citep{aharonian2007pks,albert2007mrk501,aleksic2014ic310}, and they have detected
gamma-ray bursts (GRBs) \citep{magic2019grb,hess2019grb,hess2021grb} and the
recurrent nova RS~Ophiuchi \citep{hess2022rsoph,magic2022rsoph}.

This strategy has, however, also created a blind spot: current telescopes can
detect very faint sources, but they cannot watch them continuously. This blind spot is about to matter more than ever, because the number
of transient alerts is growing by orders of magnitude. The NSF--DOE Vera C. Rubin
Observatory alone is expected to issue of order $10^{7}$ alerts per night
\citep{ivezic2019}. SKA and DSA-2000 in the radio, Einstein Probe and SVOM in
X-rays, and IceCube-Gen2 in neutrinos will add their own streams. The
gravitational-wave network is moving towards the Einstein Telescope
\citep{punturo2010}, which will issue alerts minutes before binary neutron star
mergers \citep{banerjee2023}. Many of the most interesting transients last only minutes
to days and emit a large fraction of their VHE power below 1\,TeV. This is the
case for GRBs, novae, flaring AGN, and perhaps also for some of the counterparts of
neutrinos and gravitational waves (Sect.~\ref{sec:transients}).

At the same time, the number of places from which these transients can be
observed with good sensitivity below 1\,TeV is shrinking. MAGIC, H.E.S.S.\ and
VERITAS are likely to stop operating within the next decade. The ASTRI
Mini-Array on Tenerife \citep{scuderi2022} is designed for the multi-TeV range.
The LACT array of 32 telescopes of 6\,m, planned at the LHAASO site, is designed
mainly for the ultra-high-energy range, in combination with the water Cherenkov and
muon detectors of LHAASO \citep{zhang2025lact}. Ground particle arrays such as
LHAASO, and in the future the Southern Wide-field Gamma-ray Observatory
\citep[SWGO;][]{albert2019swgo}, survey the overhead sky continuously. However, for
point sources below $\sim$1\,TeV and on timescales of minutes to days, their
sensitivity is orders of magnitude worse than that of an IACT \citep{sitarek2022}. Only
an exceptional event overcomes this limitation. LHAASO detected more than
64\,000 photons above 0.2\,TeV from GRB\,221009A, the brightest GRB ever recorded
\citep{lhaaso2023grb}, but such a burst is expected only about once every
10\,000 years \citep{burns2023boat}. By the mid-2030s, the only
IACTs with good sensitivity down to $\sim$0.1\,TeV may therefore be those at the
two CTAO sites: CTAO-North on La Palma and CTAO-South in the Atacama Desert. The
sub-TeV transient sky would then be watched from only two locations. A transient that is
not above the horizon of one of them during darkness is, in practice, not
observed at sub-TeV energies.

The solution is simple in principle: many telescopes in different time zones,
spread in longitude over both hemispheres, which observe a source one after another
as the Earth turns.
Optical time-domain astronomy has already taken this route. The Las Cumbres
Observatory operates a global network of robotic telescopes \citep{brown2013}. A
white paper submitted to the Expanding Horizons call of ESO argues that
multi-messenger astronomy in the 2040s needs a dedicated large-aperture facility
for time-domain spectroscopy, with flexible operations across both hemispheres
\citep{nissanke2025}. One proposal of this kind is the Time Domain Telescope (TDT).
It would consist of about a hundred robotic 2\,m telescopes, which together collect
as much light as one 20\,m telescope, at a much lower cost. They would be grouped in
clusters, possibly at several sites in both hemispheres, so as to obtain spectra
`anywhere, anytime' \citep{kupfer2025}. For VHE gamma rays, such an array has long
been considered unaffordable, because many telescopes can be built only
if each of them is inexpensive. We call each telescope of the array a station. An
inexpensive station is a small stand-alone telescope, which observes in
monoscopic (mono) mode (Sect.~\ref{sec:compact}). The
standard analysis of IACT data reduces each shower image to a few image parameters
\citep{hillas1985}, from which machine-learning methods such as random forests
estimate the type, the energy and the direction of the primary particle
\citep[e.g.][for LST-1]{lopezcoto2022}. With this analysis, however, a stand-alone
telescope with a 5\,m diameter mirror has a poor sensitivity and a high energy threshold,
0.73\,TeV with the photodetectors used in LST-1. With few photons from each shower and
a single viewing direction, the few image parameters carry too little information
\citep{jobst2026,paper2}. Until recently, such a telescope was therefore not
considered a useful instrument below 1\,TeV.

The analysis of \citet[][hereafter Paper~1]{jobst2026} changes this, at least in
simulation. In this analysis, which we call the transformer analysis, a video vision
transformer \citep{vaswani2017,dosovitskiy2021,arnab2021} processes the full waveform movie of a
Cherenkov camera. It therefore uses information that the few image parameters of
the standard analysis discard. The companion paper \citep[][hereafter Paper~2]{paper2}
applies it to a stand-alone 5\,m telescope with the same photomultipliers as LST-1. On
identical simulated events, the transformer analysis lowers the energy threshold by
a factor of four compared with the standard analysis, and between 0.4 and 0.63\,TeV
it is 17 times more sensitive (Sect.~\ref{sec:station}). In simulation, the 5\,m
telescope analysed with the transformer comes within a factor of about four of the
sensitivity of the 23\,m LST-1 and of MAGIC above 0.25\,TeV, although its mirror area
is 21 times smaller than that of LST-1. Compact stand-alone telescopes can
therefore be sensitive instruments below 1\,TeV, and they are inexpensive enough to
build many of them. In this paper, we argue
that the gap in temporal coverage can now be closed, and we quantify by how much. We call
the resulting concept RevolIACT: REal-sky Video-\hspace{0pt}vision-\hspace{0pt}transformer-\hspace{0pt}Optimized
Longitudinally distributed Imaging Atmospheric Cherenkov Telescopes. Its stations
are AI-driven: a transformer analyses their data, AI agents can schedule and
coordinate their observations (Sect.~\ref{sec:robotic}), and AI may also guide their
design (Paper~2).

In Sect.~\ref{sec:problem} we describe the problem of temporal coverage. In
Sect.~\ref{sec:cases} we explain why the stations are compact stand-alone telescopes,
examine which sensitivity would be sufficient for each of them, and summarise the
sensitivity that such a station reaches in simulation. In Sect.~\ref{sec:array} we
present the array and its temporal coverage, and in Sect.~\ref{sec:grb} we test it on
\GbmYears{} years of real GRBs. In Sects.~\ref{sec:feasibility} and \ref{sec:limits}
we discuss the feasibility of the array and the drawbacks of stand-alone operation,
and we conclude in Sect.~\ref{sec:conclusions}. In Appendix~\ref{app:sites} we
discuss the choice of the station sites, and in Appendix~\ref{app:zenith} we give
the results for other zenith-angle limits.

\section{The problem of temporal coverage}
\label{sec:problem}

\subsection{What one or two sites can observe}
\label{sec:twosites}

From any one site, a given source rises, reaches its highest point in the sky, and
sets. Its zenith angle,
the angle between the direction to the source and the vertical, therefore changes
continuously. We require astronomical darkness, that is, the Sun more than
$18^{\circ}$ below the horizon. Unless stated otherwise, we also require a zenith
angle below $60^{\circ}$, a typical limit for IACT observations. Section~\ref{sec:zenith65} and
Appendix~\ref{app:zenith} give the results for other limits, from $30^{\circ}$ to
$65^{\circ}$. A source on the
celestial equator is then observable from La Palma for 11.6\,\% of the year, an
average of 2.8\,h per day. We call the fraction of time during which a source can be
observed its temporal coverage, or simply its coverage. If we also require that the
Moon be below the horizon or less than 40\,\% illuminated \citep{ohm2023}, the source
is observable from La Palma for only 6.7\,\% of the year, an average of 1.6\,h per day. This time comes in windows
of a few hours, separated by gaps of most of a day. Bad weather removes a further
fraction \citep{gaug2024}. These gaps in temporal coverage cannot be filled by a
more sensitive instrument.

Two sites do not change this picture qualitatively. CTAO-North and CTAO-South are
$52^{\circ}$ apart in longitude. This leaves $308^{\circ}$ of longitude, more than
twenty hours of Earth rotation, without a CTAO site. Averaged over the sky and the
year, a source is observable from at least one of the two sites for 17\,\% of the
time (Sect.~\ref{sec:coverage}). Closing this gap requires many sites, spread in
longitude over both hemispheres.

The limited coverage does not only leave gaps in the light curves: it decides what
can be measured at all. Many of the processes at work in VHE sources, such as the acceleration and cooling of
the particles and the evolution of the region that emits, produce changes that we
observe on timescales of hours to days. Models of these processes are tested with light curves
that follow the source through these changes in every band, from radio to TeV
energies. Multi-day monitoring with simultaneous multiwavelength coverage is
routine in the radio, optical and X-ray bands and at GeV energies, but not at
sub-TeV energies. When the sub-TeV band is missing for most of each day, the
interpretation must rely on assumptions about what happened in the gaps.

Three recent studies of the blazar Mrk~421 show what a denser VHE coverage
enables. The light curves of these studies are not continuous: each site observed the
source for a few hours per night. These light curves have, however, far more VHE
coverage than most, thanks to
the dense sampling of MAGIC, complemented in two of the studies by VERITAS, six hours
of Earth rotation further west. In April 2013, an outburst was followed on nine
consecutive nights, with 43 hours of simultaneous VHE and X-ray observations. The
VHE flux reached about 15 times that of the Crab Nebula, the reference source of
VHE astronomy, and the spectrum extended beyond 10\,TeV. The data
gave 240 broadband spectral energy distributions of 15 minutes each. They show
several spectral hysteresis loops in X-rays, and the first evidence of spectral
hysteresis at VHE, simultaneous with the X-ray loops. A time-dependent model
attributes the sub-hour variability to changes in the luminosity and spectral slope
of the freshly accelerated electrons, with a nearly constant magnetic field and size
of the emitting region. The model therefore favours acceleration at a shock over
magnetic reconnection \citep{magic2026mrk421}. A campaign from November 2009 to July 2010
followed the source through its most violent period, which included the
highest VHE flux ever measured from an AGN. The X-ray and VHE fluxes were the most
variable, and they were strongly correlated, without a time lag.
The broadband data are consistent with a compact emitting zone embedded in a larger
one \citep{schmuckermaier2025}. Within this period, MAGIC observed a flare on sixteen
nights between 8 and 26 January 2010. On three of them, the spectra show the electron
pile-up predicted by stochastic acceleration, at the transition from
acceleration-dominated to cooling-dominated evolution \citep{abhir2026}. Unfortunately, such
datasets remain rare, because VHE observing time is scarce.

Many such campaigns on AGN and Galactic binary systems are coordinated with
expensive and heavily oversubscribed facilities, such as \textit{Chandra},
\textit{XMM-Newton}, \textit{NuSTAR}, IXPE and the Event Horizon Telescope
\citep[e.g.][]{anderhub2009,algaba2021,magic2025ixpe}. Their
observing time is allocated once, often months in advance. If the weather is bad
at the site of the IACT on that night, the sub-TeV band is lost for good. The
expensive radio or X-ray observation has then been made without simultaneous
sub-TeV observations. It cannot simply be repeated. A new observation would need
a new allocation, months later, and by then the source would be in a different
state, because no two flares are the same. The loss is largest for two kinds of events, for which a missed
observation can never be repeated. The first are fast transients, such as GRBs and
perhaps the counterparts of some neutrino and gravitational-wave alerts. The VHE flux
of a GRB is highest in the first minutes and then falls quickly
(Sect.~\ref{sec:sensitivity}). If no telescope can observe the burst during these
minutes, its brightest phase is lost, even if the burst is observed a few hours later.
The second are singular events (Sect.~\ref{sec:singular}), such as the neutron star
merger GW\,170817. Such events are rare, and each of them happens only once. If no
telescope can observe such an event when it occurs, its sub-TeV emission is never
measured.

\subsection{Time-domain astronomy and the growing number of transients}
\label{sec:transients}

This limitation will matter more in the coming decade, because time-domain and
multi-messenger astronomy are entering a new era. Wide-field instruments now
survey the sky at almost every wavelength, every few nights in the optical and
continuously in other bands, and they detect transient sources in large numbers.
In addition, neutrino and gravitational-wave observatories now issue alerts in real time.
Table~\ref{tab:alerts} lists the main facilities and the rates of their alerts or
detections.

\begin{table*}[t]
\caption{Main wide-field facilities that detect transients or issue alerts, with the
rates of their alerts or detections. Many of these alerts concern sources that are
expected to emit gamma rays below 1\,TeV, although they are a small fraction of all
alerts (Sect.~\ref{sec:transients}).}
\label{tab:alerts}
\centering
\footnotesize\setlength{\tabcolsep}{4pt}
\renewcommand{\arraystretch}{1.1}
\begin{tabular}{@{}>{\raggedright}p{0.20\textwidth} >{\raggedright}p{0.10\textwidth}
                >{\raggedright}p{0.22\textwidth} >{\raggedright}p{0.20\textwidth}
                >{\raggedright\arraybackslash}p{0.20\textwidth}@{}}
\toprule
Facility & Band or messenger & Alerts or detections & Rate & References \\
\midrule
Zwicky Transient Facility (since 2018) & Optical & Transient alerts
  & Up to $1.2\times10^{6}$ per night & \citealt{bellm2019,patterson2019} \\
Vera C. Rubin Observatory, Legacy Survey of Space and Time (from 2026, for ten years)
  & Optical & Transient alerts & Of order $10^{7}$ per night
  & \citealt{ivezic2019,findeisen2026} \\
ULTRASAT (planned) & Ultraviolet & Transients in large parts of the sky & --
  & \citealt{shvartzvald2024} \\
CHIME/FRB & Radio & Fast radio bursts (FRBs) & About 900 per year (2018--2023)
  & \citealt{chime2026} \\
CHORD, DSA-2000, SKA (early 2030s) & Radio & New FRB sources
  & Tens of thousands per year & \citealt{hallinan2019,dewdney2009,curtin2026} \\
Einstein Probe, SVOM (launched in 2024) & X-rays, gamma rays
  & X-ray and gamma-ray transients & Hundreds per year, together with \textit{Fermi}
  & \citealt{yuan2025,wei2016} \\
\textit{Fermi}, GBM and LAT (since 2008) & Gamma rays
  & GRBs and other gamma-ray transients
  & GBM alone: \GbmPerYear{} GRBs per year (Sect.~\ref{sec:grbmethod})
  & \citealt{meegan2009,atwood2009} \\
IceCube & Neutrinos & Single high-energy neutrinos, alerts within minutes
  & About 30 per year & \citealt{aartsen2017rt,abbasi2023icecat} \\
IceCube-Gen2 (planned) & Neutrinos & Astrophysical neutrinos
  & Ten times the rate of IceCube & \citealt{aartsen2021gen2} \\
KM3NeT (under construction) & Neutrinos & High-energy neutrinos & --
  & \citealt{adrian2016} \\
P-ONE, TRIDENT, HUNT (in development) & Neutrinos & High-energy neutrinos
  & Not yet estimated; volumes from about 1\,km$^{3}$ (P-ONE) to 30\,km$^{3}$ (HUNT),
    compared with 1\,km$^{3}$ for IceCube
  & \citealt{agostini2020pone,ye2023trident,huang2023hunt} \\
Hyper-Kamiokande (from 2028) & Low-energy neutrinos
  & Neutrino bursts of core-collapse supernovae in the Galaxy and up to the Andromeda
    galaxy, before the light of the explosion; direction to about $1^{\circ}$ at 10\,kpc
  & About 1.6 Galactic core-collapse supernovae per century
  & \citealt{hyperk2018,bian2022hk,alkharusi2021snews,rozwadowska2021} \\
LIGO, Virgo, KAGRA, fifth observing run (O5) & Gravitational waves
  & Binary mergers, mostly of black holes; alerts within minutes
  & Several hundred to about a thousand per year
  & \citealt{abbott2020lrr,kiendrebeogo2023} \\
Einstein Telescope, Cosmic Explorer (next generation) & Gravitational waves
  & Binary mergers out to cosmological distances; some neutron star mergers announced
    minutes in advance
  & Einstein Telescope: $10^{5}$--$10^{6}$ binary black hole and $7\times10^{4}$
    binary neutron star mergers per year
  & \citealt{punturo2010,reitze2019,maggiore2020,banerjee2023} \\
\bottomrule
\end{tabular}
\end{table*}

These alerts include many sources that are expected to emit
gamma rays below 1\,TeV: GRBs, including those produced by neutron star mergers,
novae, flaring AGN, and the sources of high-energy neutrinos
\citep[e.g.][]{cta2019}. They are a small fraction of all alerts, but a large number.
Their VHE emission
typically lasts minutes to days, and for GRBs the first minutes are the brightest.
To measure this emission, a telescope must point at the source within minutes of
the alert, or must happen to be observing that part of the sky. A non-detection is
also useful, because an upper limit on the sub-TeV flux constrains the emission
models. But if no telescope is observing, nothing is learnt about the sub-TeV
emission: we do not know whether sub-TeV gamma rays were emitted or not. A telescope
can also point at only one region of the sky at a time. With such high alert rates, one
must choose which alerts to follow, even among the sources that may emit below
1\,TeV (Sect.~\ref{sec:triage}). Next-generation radio, optical, X-ray,
neutrino and gravitational-wave facilities will detect orders of magnitude more such events. Without corresponding coverage in the sub-TeV band,
much of their joint multi-messenger potential will remain unrealised.

\subsection{Singular events: no second chance}
\label{sec:singular}

Singular events can teach us a great deal about the Universe, and can rewrite entire
fields within days. On 23 February 1987, underground detectors recorded about two
dozen neutrinos from SN\,1987A within 13\,s
\citep{hirata1987,bionta1987,arnett1989}. These neutrinos confirmed that massive
stars end their lives in the collapse of their core \citep{arnett1989}. Within
weeks, they gave a limit on the mass of the neutrino, and later limits on new
particles such as axions \citep{arnett1989,raffelt1990}. They also opened neutrino
astronomy beyond the Solar System, which was recognised by the Nobel Prize in
Physics in 2002. Thirty years after SN\,1987A, the gamma-ray burst GRB\,170817A arrived
1.7\,s after the gravitational waves of the neutron star merger GW\,170817
\citep{abbott2017}. This delay showed that the speeds of gravitational waves and of
light differ by at most a few parts in $10^{15}$. Within a day of the announcement, this
result ruled out many theories of dark energy \citep{ezquiaga2017}. The kilonova
that followed showed that such mergers forge heavy elements \citep{pian2017}, and
the event gave a new way to measure the expansion rate of the Universe
\citep{abbott2017h0}.

Sub-TeV gamma rays contributed the least to both events. In 1987, the imaging
Cherenkov technique was still two years away from its first detection
\citep{weekes1989}. In 2017, GW\,170817 occurred in broad daylight at the MAGIC
site, and no IACT could observe it until hours later. H.E.S.S.\ started to observe it
5.3\,h after the merger, at a zenith angle of about $60^{\circ}$, which raised its
energy threshold to 0.28\,TeV \citep{hess2017gw}. These conditions were far from
optimal for constraining the emission, and no gamma rays were detected. Even the deep
upper limits obtained months later, around the peak of the afterglow
\citep{hess2020gw}, lie well above the flux that a model of the afterglow predicts
for that time \citep{pellouin2024}. GW\,170817 remains the only gravitational-wave event
with an electromagnetic counterpart. MAGIC found a hint of a signal, at about
$3\sigma$, from the nearby short GRB\,160821B, which probably also came from a
neutron-star merger \citep{acciari2021grb}. Whether such mergers accelerate particles
to TeV energies in the first hours is still unknown. The next singular event will
arrive unannounced. Whether we can measure it at sub-TeV energies will again depend
on which instruments are observing when it occurs.

\section{Compact stand-alone stations}
\label{sec:cases}

Sensitivity sets the time resolution of a light curve, coverage sets its length and
continuity, and both are needed. Many stations are affordable only if each of them
is inexpensive, and they are useful only if each of them is sensitive enough to detect the
transients that it observes. When a source is not detected, the sensitivity must
still be sufficient for the upper limit on its flux to constrain the emission
models. In this section we explain why an inexpensive station is a compact stand-alone
telescope (Sect.~\ref{sec:compact}), examine which sensitivity would be sufficient for such
a station (Sect.~\ref{sec:sensitivity}), and compare it with the sensitivity that the
station reaches in a first, simplified and idealised simulation
(Sect.~\ref{sec:station}).

\subsection{Why compact and stand-alone}
\label{sec:compact}

Closing the gap in coverage requires many stations spread in longitude
(Sect.~\ref{sec:twosites}). The array of Sect.~\ref{sec:array} has ten to fourteen
of them. An array of this size is affordable only if each station is inexpensive. The
cost of the dish and of the telescope structure rises faster than the square of the dish diameter. Stereoscopic observation multiplies the cost of
a site by the number of telescopes that see each shower, and adds an array trigger.
An inexpensive station is therefore a single compact telescope, which observes in mono
mode. A
compact telescope can have some advantages over a large one, which may be
particularly useful for building an array at limited cost. Examples are a
retractable enclosure, fast repointing, robotic operation and a low-cost camera
with a large field of view (Sect.~\ref{sec:array}). The drawbacks of this choice
are that fewer Cherenkov photons are detected from each shower, and that each
shower is seen from only one direction (Sect.~\ref{sec:station}).

\subsection{What sensitivity would be sufficient}
\label{sec:sensitivity}

In the last decade we have learnt that many transients, with timescales of
minutes to days, are detected above 0.1--0.2\,TeV with fluxes of order 0.1 of the
flux of the Crab Nebula (Crab units, C.U.) or more. We take the flux of the Crab
Nebula from the log-parabola measured by MAGIC \citep[Eq.~1]{aleksic2016}:
$5.5\times10^{-10}$\,cm$^{-2}$\,s$^{-1}$ above 0.1\,TeV and
$2.3\times10^{-10}$\,cm$^{-2}$\,s$^{-1}$ above 0.2\,TeV. The VHE flux of
GRB\,190114C exceeded 100\,C.U.\ in the first minutes after the burst and
decreased by about two orders of magnitude within half an hour
\citep{magic2019grb}. Other GRBs detected at VHE show that the emission can last
from tens of minutes to more than two days
\citep{hess2019grb,hess2021grb,magic2024grb}. In the bursts with the
best-measured light curves, the VHE flux falls rapidly with time, approximately as
a power law $F\propto t^{-\alpha}$ with $\alpha$ between about 1.1 and 1.6
\citep{magic2019grb,hess2021grb}. Bursts like the nearby GRB\,190829A, at a
redshift of 0.08, are among the main targets of the array. This burst was weak: assuming
isotropic emission, it released only about $2\times10^{50}$\,erg in gamma rays during
the burst \citep{hess2021grb}, about a thousand times less than GRB\,190114C
\citep{ravasio2019}. However, because it was nearby, its afterglow was bright
at VHE: H.E.S.S. observed it on three nights, starting 4.3\,h after the burst, and
detected it on the first two \citep{hess2021grb}. Its flux above 0.2\,TeV was about
0.11\,C.U.\ on the first night and 0.02\,C.U.\ on the second, and it decreased with
time as $t^{-1.09}$ \citep{hess2021grb}. With this law, the mean flux of the first
night, observed from 4.3 to 7.9\,h after the burst, equals the flux 5.9\,h after
the burst. One day after the burst, the flux would therefore have been
$0.11\times(24/5.9)^{-1.09}\approx0.024$\,C.U., and the same law reproduces the
flux of the second night. Such nearby bursts are rare but not unique: GRB\,171205A,
at a redshift of 0.037 \citep{izzo2019}, and GRB\,211211A, at 350\,Mpc
\citep{rastinejad2022}, were also closer than a redshift of 0.1. Weak but nearby
bursts like GRB\,190829A will occur again.
Following them without interruption, from the trigger into the second day, requires
stations distributed in longitude, such as those of RevolIACT.

The recurrent nova RS~Ophiuchi was detected for several days after its eruption in
August 2021 \citep{hess2022rsoph,magic2022rsoph}. Over the first four days, its
flux above 0.1\,TeV was $4.4\times10^{-11}$\,cm$^{-2}$\,s$^{-1}$, or 0.08\,C.U.\
\citep{magic2022rsoph}. Its spectrum was steep, so that its flux above 0.2\,TeV
was lower in Crab units. For this flux we use the spectra that H.E.S.S. measured
night by night \citep{hess2022rsoph}, because they have the widest spectral
coverage above 0.2\,TeV. From the H.E.S.S. spectra of the first five nights,
we derive $(0.6$--$2.3)\times10^{-11}$\,cm$^{-2}$\,s$^{-1}$ above 0.2\,TeV, or
0.03--0.10\,C.U., and 0.06\,C.U.\ on average. The blazar
TXS\,0506+056 was detected at VHE after the high-energy neutrino IceCube-170922A
\citep{icecube2018,ansoldi2018}. On its two brightest nights, 3 and 4 October 2017,
its spectrum was a steep power law, with a photon index of about 3.9
\citep{ansoldi2018}. Its flux was then about 0.1\,C.U.\ above 0.09\,TeV and
$4.4\times10^{-11}$\,cm$^{-2}$\,s$^{-1}$, or 0.08\,C.U., above 0.1\,TeV, but only
$6.0\times10^{-12}$\,cm$^{-2}$\,s$^{-1}$, or 0.03\,C.U., above 0.2\,TeV. Flares of
AGN often exceed 0.2\,C.U.\ and sometimes 1\,C.U.\ \citep[e.g.][]{acciari2020}.
The brightest VHE blazars, such as Mrk~421, Mrk~501, 1ES\,1959+650, 1ES\,2344+514
and PG\,1553+113, are the targets of long-term multi-instrument campaigns
\citep[e.g.][]{magic2021mrk421,magic2023mrk501,magic2024es2344,magic2024pg1553}. A
flux of 0.2\,C.U.\ is typical of these sources outside their strongest flares, and
such a flux is easily reached by moderate flares of fainter AGN. The detection time
of a source of 0.2\,C.U.\ therefore measures the ability of a single station to monitor the brightest AGN in their low
states, and moderate flares of fainter AGN.

The sensitivity of an instrument is the weakest flux that it can detect in a given
observing time with a statistical significance of five standard deviations ($5\sigma$).
When we compute sensitivities (Sect.~\ref{sec:station}), we use the standard
criteria: a $5\sigma$ excess according to Eq.~(17) of \citet{lima1983}, with the
background estimated from five OFF regions, at least ten excess events ($N_{\rm ex}\geq10$), and an
excess of at least 5\,\% of the background ($N_{\rm ex}\geq0.05\,N_{\rm bkg}$;
\citealt{aleksic2016}), where $N_{\rm ex}$ and $N_{\rm bkg}$ are the numbers of excess
and background events. The current major
IACT systems, with two to five telescopes observing in stereo, reach an integral point-source sensitivity of about 1\,\%~C.U.\ in 25\,h above 0.2--0.3\,TeV,
equivalent to roughly $\sens{5}$ \citep{aleksic2016,holder2011,aharonian2006}, that is,
about 5\,\%~C.U.\ in one hour, which improves as the square root of the observing time
$T$ in hours. The exact value differs between instruments, and it depends on the
definition of the sensitivity (Sect.~\ref{sec:station}). CTAO
will improve on this by a factor of several \citep{cta2019}. In the following, we
distinguish the sensitivity of such an IACT system from that of one station of the
array, and from that of the array, in which the data of several stations are
combined. In the background-dominated regime, an instrument whose integral sensitivity in one hour is $S$ detects a source of flux $F$ at $5\sigma$ in an
observing time
\begin{equation}
T \;=\; \left(\frac{S}{F}\right)^{2}\,\mathrm{h}.
\label{eq:tdet}
\end{equation}
The observing time falls with the square of the source flux and grows with the
square of $S$. For very bright sources, Eq.~(\ref{eq:tdet}) gives very short
observing times, and the requirement of at least ten excess events sets the
detection time instead.

Equation~(\ref{eq:tdet}) is only an approximation. The sensitivity $S$ refers to a
source with the spectrum of the Crab Nebula. For a source with a different
spectrum, the observing time needed for a detection differs from that given by
Eq.~(\ref{eq:tdet}). A steeper spectrum, for example, puts more of the flux close
to the energy threshold, where a smaller fraction of the gamma rays is detected,
and therefore needs a longer observing time. We use Eq.~(\ref{eq:tdet}) not for accurate
predictions, but as a simple way to make a rough estimate of the observing time,
which puts the fluxes of the transients and the sensitivities of the instruments
into context.

\begin{table}[t]
\caption{Approximate observing time needed for a $5\sigma$ detection of several
important VHE transients of the last decade, from Eq.~(\ref{eq:tdet}), for a
typical integral sensitivity of the current major IACT systems (MAGIC, H.E.S.S.\ and
VERITAS) above 0.2\,TeV,
$S=\sens{5}$, and for sensitivities per station three and nine times worse,
$S=\sens{15}$ and $\sens{45}$. The fluxes are those above 0.2\,TeV, and the
sensitivity is defined in Sect.~\ref{sec:sensitivity}.}
\label{tab:cases}
\centering
\small\setlength{\tabcolsep}{2.7pt}
\begin{tabular}{@{}l c r r r@{}}
\toprule
Transient & Flux          & \multicolumn{3}{c}{$T$ for $5\sigma$} \\
          & $>$0.2\,TeV   & \multicolumn{3}{c}{at $S=$} \\
\cmidrule(lr){3-5}
          & (C.U.)        & 5\,\% & 15\,\% & 45\,\% \\
\midrule
GRB\,190114C afterglow,\tablefootmark{a}       & $>10$ & few s\tablefootmark{g}
                                                       & few s\tablefootmark{g}
                                                       & $\lesssim$10\,s\tablefootmark{g} \\
\quad first minutes                            &       &          &          &         \\
Bright AGN flare\tablefootmark{b}              & 1     & 44\,s\tablefootmark{h}
                                                       & 80\,s    & 12\,min \\
Bright AGN in low state,\tablefootmark{c}      & 0.2   & 3.8\,min
                                                       & 34\,min  & 5.1\,h  \\
\quad or moderate flare                        &       &          &          &         \\
\quad of a fainter AGN                         &       &          &          &         \\
GRB\,190829A afterglow,\tablefootmark{d}       & 0.11  & 12\,min  & 1.9\,h   & 17\,h   \\
\quad first night                              &       &          &          &         \\
Nova RS~Ophiuchi (Aug.\ 2021),\tablefootmark{e} & 0.06 & 42\,min  & 6.3\,h   & 56\,h   \\
\quad sustained for days                       &       &          &          &         \\
TXS\,0506+056 (Sep.--Oct.\ 2017)\tablefootmark{f} & 0.03 & 2.8\,h & 25\,h    & 225\,h  \\
\quad (neutrino counterpart)                   &       &          &          &         \\
GRB\,190829A afterglow,\tablefootmark{d}       & 0.024 & 4.3\,h   & 39\,h    & 352\,h  \\
\quad one day after the burst                  &       &          &          &         \\
\bottomrule
\end{tabular}
\tablefoot{
\tablefoottext{a}{\citet{magic2019grb}.}
\tablefoottext{b}{\citet{acciari2020}.}
\tablefoottext{c}{Proxy for the monitoring of the brightest VHE blazars in their low
states \citep[e.g.][]{magic2021mrk421,magic2023mrk501,magic2024es2344,magic2024pg1553},
and of moderate flares of fainter AGN.}
\tablefoottext{d}{\citet{hess2021grb}; observed spectrum of the first night, 4.3 to
7.9\,h after the burst, at a redshift of 0.08; 0.02\,C.U.\ on the second night. One
day after the burst: flux of the first night scaled with the measured decay,
$t^{-1.09}$ (see text).}
\tablefoottext{e}{Average over the first five nights, from the nightly spectra of
\citet{hess2022rsoph}; 0.03--0.10\,C.U.\ from night to night.}
\tablefoottext{f}{\citet{icecube2018,ansoldi2018}; brightest nights, 3 and 4
October 2017, after the neutrino of 22 September.}
\tablefoottext{g}{Eq.~(\ref{eq:tdet}) gives 0.1, 1 and 7\,s. At such short
observing times the minimum of ten excess events dominates: for an effective area of
about $10^{5}$\,m$^{2}$, that of the SVT with the transformer analysis above
0.25\,TeV (from the rates of Fig.~\ref{fig:sens}), a source of 10\,C.U.\ yields ten gamma rays above
0.2\,TeV in about 4\,s.}
\tablefoottext{h}{Eq.~(\ref{eq:tdet}) gives 9\,s. For the same effective area,
ten gamma rays above 0.2\,TeV take about 44\,s at 1\,C.U., which sets the
observing time.}}
\end{table}

To find which sensitivity would be sufficient, we evaluate Eq.~(\ref{eq:tdet}) in
Table~\ref{tab:cases} for the transients described above, which are among the most
important VHE transients detected during the last decade. Table~\ref{tab:cases}
gives these observing times for three sensitivities. The first, $\sens{5}$, is
approximately that of the current major IACT systems. The other two, $\sens{15}$ and
$\sens{45}$, are three and nine times worse, and are possible sensitivities of one
station.
The current IACT systems detect the first five transients within an hour, the
neutrino counterpart TXS\,0506+056 within three hours, and GRB\,190829A one day after
the burst within about four hours.

A station three times less sensitive, at $\sens{15}$, still detects a GRB afterglow
in its first minutes within seconds, a bright AGN flare within two minutes, and a
source of 0.2\,C.U.\ within about half an hour. A single station can therefore
monitor the brightest VHE blazars even in their low states, and follow moderate
flares of fainter AGN, with good time resolution. The First G-APD Cherenkov Telescope
(FACT), a stand-alone telescope with a mirror area of 9.5\,m$^{2}$, has monitored the
brightest blazars at TeV energies in this way for years
\citep{anderhub2013,arbetengels2021}. A station can also issue alerts on the flares
that it detects, so that CTAO can observe them in depth
(Sect.~\ref{sec:sensperstation}). The afterglow of GRB\,190829A takes about two
hours, within the first night of observation. One day after the burst, it takes 39
hours, which the array collects in about a day by combining the stations' observations
(see below). The array could have observed this burst without interruption for two days
(Sect.~\ref{sec:grbresults}). Since the afterglow faded with time, the array could
thus have detected it continuously from the trigger until about one day after the
burst. A nova like RS~Ophiuchi takes about six hours, which one site can collect
only in a long night. The array detects it within a few hours. A neutrino counterpart
like TXS\,0506+056 takes 25 hours, far more than one night at one site, but the array
collects this observing time within a day.

A station about nine times less sensitive than the current IACT systems, at $\sens{45}$,
needs nine times longer than one at $\sens{15}$. For bright flares this costs time
resolution: a light curve binned at 12 minutes instead of 80 seconds. For
GRB\,190829A, the nova and the neutrino counterpart, this factor is decisive. The 17 hours
needed for GRB\,190829A exceed its first night, and one day after the burst the
array would need about nine days. The 56 hours needed for the nova take
even the array more than a day, so that its evolution within a day could not be
followed. The 225 hours needed for TXS\,0506+056 take the array about six days,
longer than its flare lasted.

An array of stations, each at $\sens{15}$, adds a further capability, because
several stations observe the same source within a day and their data can be
combined. For a source on the celestial equator, on average 1.6 of the fourteen
stations of Sect.~\ref{sec:array} observe it in darkness (Sect.~\ref{sec:coverage}),
which adds up to about 38 station-hours per day. Combined, they give an array
sensitivity of about 2.4\,\%~C.U.\ in one day (Eq.~\ref{eq:tdet}), slightly better
than the 3\,\%~C.U.\ that a current IACT system on La Palma (e.g. MAGIC) reaches in
the 2.8\,h per day during which it can observe such a source. The array can
therefore detect transients well below 0.1\,C.U.\ if they last a day or more. This
estimate neglects the higher energy threshold of the stations that observe at large
zenith angles. For MAGIC, the threshold grows approximately as
$(\cos\theta_{\rm z})^{-2.3}$, where $\theta_{\rm z}$ is the zenith angle
\citep{aleksic2016}. A station with a threshold of 0.075\,TeV near the zenith would therefore reach about
0.2\,TeV at a zenith angle of $50^{\circ}$. For the 5\,m telescope near the zenith, Paper~2 finds a threshold of 0.18\,TeV with
the photomultipliers of LST-1, and 0.11\,TeV with photodetectors of 50\,\% quantum
efficiency (QE). With the photomultipliers of LST-1, a 6\,m dish gives 0.13\,TeV. Better photodetectors and a better
analysis should bring the threshold of a station towards 0.075\,TeV, which we take as
the goal (Sect.~\ref{sec:station}). Up to this angle, its sensitivity above 0.2\,TeV
should change little, because at the energies above the threshold the collection
area grows with the zenith angle, as the Cherenkov light pool becomes larger
\citep{magic2020vlza}. With a limit of $50^{\circ}$, on average 1.2 of the fourteen
stations observe a source on the celestial equator, or 29 station-hours per day.
Combined, they reach about 2.8\,\%~C.U.\ in one day, still better than the
3.4\,\%~C.U.\ of a current IACT system on La Palma with the same limit. The
conclusion therefore holds when the zenith angle is restricted so that the threshold
stays near 0.2\,TeV. Table~\ref{tab:zenith} gives these numbers for every
zenith-angle limit of Appendix~\ref{app:zenith}.

\paragraph{A low energy threshold adds further science.} Good sensitivity below 0.1\,TeV, and
even below 0.05\,TeV, gives access to transients at large distances, such as distant
GRBs and AGN. The extragalactic background light absorbs the VHE photons of sources
at cosmological distances, but fewer of them at the lowest energies
\citep[e.g.][]{dominguez2011}. The two GRBs that
MAGIC detected within a minute of the burst lie at redshifts of 0.42 and 1.1
\citep{magic2019grb,magic2024grb}. The CTAO LST sub-array will excel in this energy range
\citep{cta2019}. For the stations, a useful sensitivity down to $\sim$0.1\,TeV is
desirable, although it is not required for the transients of Table~\ref{tab:cases}.
Near the zenith, such a sensitivity is possible. Below $30^{\circ}$, the threshold is at most 1.4 times
higher than at the zenith, or about 0.1\,TeV for a station with a threshold of
0.075\,TeV at the zenith (Table~\ref{tab:zenith}). Observing only below $30^{\circ}$
would, however, reduce the coverage of fourteen stations, averaged over the sky,
from 61\,\% to 28\,\% (Table~\ref{tab:zenith}). The limit of $60^{\circ}$ does not
remove the observations below $30^{\circ}$. It adds
those at larger zenith angles, where the threshold is higher. The low threshold,
and with it the potential to discover more distant sources, is therefore available
on average for almost half of the time during which the array observes a source.

To conclude, a flux sensitivity per station of about $\sens{15}$ above
0.2--0.3\,TeV, about three times worse than that of the current major IACT systems, would be
sufficient for the main science goals of the array. With it, a single station detects within an hour the GRB afterglow in its first
minutes, the bright AGN flares, the brightest blazars in their low states and moderate
flares of fainter AGN. It detects a weak but nearby GRB like GRB\,190829A within about
two hours of its first night, and a nova like RS~Ophiuchi within a long night. The array detects the neutrino counterpart
TXS\,0506+056 within a day, and follows the afterglow of GRB\,190829A until about
one day after the burst. A sensitivity per station about nine times worse than that of
the current IACT systems, $\sens{45}$, is not sufficient for the fainter of these transients. Ten
to fourteen stations at $\sens{15}$ each, however, can combine their data. For a
transient on the celestial equator that lasts a day, they reach an array sensitivity
of 2.4--2.9\,\%~C.U., comparable to or better than the 3\,\%~C.U.\ of a current IACT
system on La Palma. We therefore use $\sens{15}$ per station as an indicative value,
not as a strict requirement. A better sensitivity shortens every observing time, and
with a somewhat worse sensitivity a station still detects the brighter transients. A useful sensitivity
down to $\sim$0.1\,TeV is also desirable, for the more distant sources. At
$\sens{15}$ per station, a source of 0.1\,C.U.\ is detected at $5\sigma$ in about
2.3\,h, within the observing window of a single station.

\subsection{What a compact AI-driven station reaches}
\label{sec:station}

\begin{figure*}[t]
\centering
\includegraphics[width=0.49\textwidth]{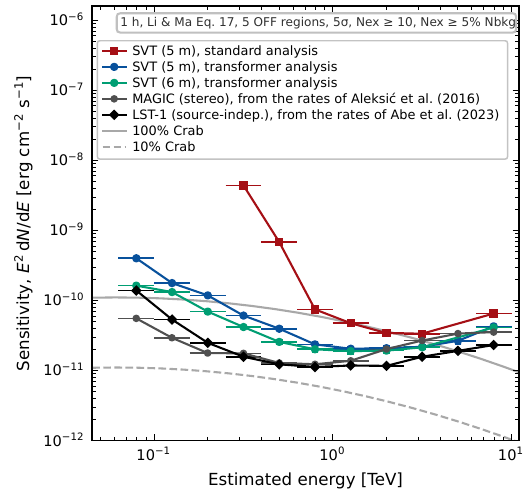}
\hfill
\includegraphics[width=0.49\textwidth]{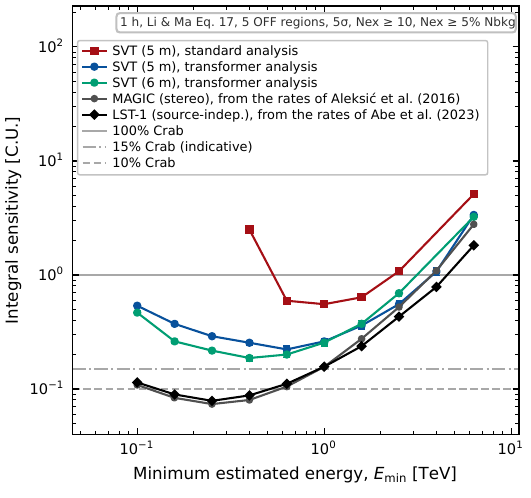}
\caption{Point-source sensitivity in one hour of the 5\,m Small Virtual Telescope
(SVT) of Papers~1 and 2, with the quantum efficiency of the LST-1 photomultipliers,
analysed with the standard analysis (red) and with the transformer analysis (blue),
and of the SVT with a 6\,m dish, analysed with the transformer analysis (green). The
sensitivities of MAGIC in stereo and of LST-1 are computed from their rates of gamma
rays and background, for one hour (Paper~2, Appendix~A). Lower curves mean better sensitivity: each point gives the weakest flux of
a source that can be detected in one hour. All curves use the definition of the
sensitivity given in Sect.~\ref{sec:sensitivity}. The grey lines show 100\,\% and
10\,\% of the flux of the Crab Nebula \citep{aleksic2016}. In the right panel, the
dash-dotted grey line shows 15\,\% of the Crab flux, the indicative sensitivity per
station of Sect.~\ref{sec:sensitivity}. Left: differential
sensitivity, as $E^{2}\,\mathrm{d}N/\mathrm{d}E$, in bins of estimated energy
(horizontal bars). Right: integral sensitivity above each threshold of estimated
energy, in units of the Crab flux above that threshold (C.U.).}
\label{fig:sens}
\end{figure*}

A compact stand-alone telescope starts with two disadvantages. The effective mirror
area of the 5\,m telescope of Papers~1 and 2 is about 18\,m$^{2}$, 21 times smaller than
that of LST-1, so that it collects about twenty times fewer Cherenkov photons from
each shower. It also sees each shower from a single direction, so that the
stereoscopic reconstruction of the shower geometry is not available. With the
standard analysis, both disadvantages raise the energy threshold and degrade the
sensitivity. With the photodetectors of
LST-1, the standard analysis reaches no energy bin below 0.25\,TeV, and its energy
threshold is 0.73\,TeV (Paper~2). This is why compact telescopes have so far been designed for the
multi-TeV range \citep[e.g.][]{scuderi2022}, or, like FACT, for the monitoring of
bright sources at TeV energies \citep{anderhub2013}.

The transformer analysis of Paper~1 removes much of this disadvantage, at least in
simulation.
Figure~\ref{fig:sens} shows the sensitivity, from Paper~2, of the 5\,m telescope of
Paper~1, the
Small Virtual Telescope (SVT), with the QE of the photomultipliers installed in
LST-1. The SVT has a 5\,m parabolic dish
with a focal ratio of 1.5, a square camera of $60\times60$ square pixels,
$0.1^{\circ}$ on a side, which covers $6.1^{\circ}\times6.1^{\circ}$, a
single-photoelectron pulse of 1\,ns full width at half maximum, and waveforms sampled
at 5\,GS/s. It observes at a zenith angle of $10^{\circ}$ from the CTAO-North site on
La Palma. For simplicity, the simulated background consists of protons only, the
dominant component of the cosmic rays. Helium and heavier nuclei, and cosmic-ray
electrons, are neglected (Paper~2). The same simulated events are
analysed with the standard analysis and with the transformer analysis. For
comparison, the figure shows the sensitivity curves of MAGIC in stereo and LST-1,
computed in the same way from their rates of gamma rays and background (Paper~2). For MAGIC, these are the
rates that \citet{aleksic2016} measured at zenith angles below $30^{\circ}$. For
LST-1, they are the rates of its source-independent analysis \citep{abe2023}, with
cuts optimised for one hour (Paper~2, Appendix~A). The sensitivities of the SVT come from simulations that
are still simple and idealised, whereas those of MAGIC and LST-1 come from real
observations of the Crab Nebula. The comparison is therefore only meant to put the
SVT into context (Paper~2).

The left panel shows the differential sensitivity. In each of five bins per decade
of estimated energy, it is the lowest flux of a source with the spectrum of the Crab
Nebula that meets, in one hour, the criteria of Sect.~\ref{sec:sensitivity}. We give
the sensitivity in bins of estimated energy because an observation measures the
estimated energy, as is usual for IACTs \citep[e.g.][]{aleksic2016}. The
differential curves are those of Figs.~1 and 3 of Paper~2, which computes LST-1 and
MAGIC in the same way. With the transformer analysis, the
sensitivity extends down to the bin of 0.063--0.1\,TeV. The standard analysis reaches
no bin below 0.25\,TeV, and between 0.4 and 0.63\,TeV it is 17 times less
sensitive than the transformer analysis. Compared with MAGIC, the SVT with the
transformer analysis is 6.0--6.7 times less sensitive between 0.1 and 0.25\,TeV,
3.1--3.5 times between 0.25 and 0.63\,TeV, and 1.5--1.9 times between 0.63 and
1.6\,TeV. It is as sensitive as MAGIC between 1.6 and 2.5\,TeV, and 1.2--1.3 times
more sensitive between 2.5 and 6.3\,TeV, where the sensitivity of MAGIC in one hour
is limited by the minimum of ten excess events. At these energies, so little
background survives the cuts that the better background rejection of MAGIC, which
comes from stereoscopy, no longer determines the sensitivity. The sensitivity is
then set by the number of gamma rays detected, and so by the collection area. The
collection area of MAGIC is smaller, because each shower must trigger both
telescopes and because its camera has a smaller field of view
(Sect.~\ref{sec:array}). Between 1.6 and 6.3\,TeV, the SVT with the transformer
analysis detects 1.3--1.7 times more gamma rays from the Crab Nebula than MAGIC,
after all cuts. Compared with LST-1, the SVT is 3.2--4.8 times less sensitive between 0.1 and
0.63\,TeV, and 1.4--2.1 times between 0.63 and 6.3\,TeV. As in Paper~2, the
comparisons stop at 6.3\,TeV. The simulated gamma rays end at 10\,TeV, so that the
last bin, 6.3--10\,TeV, misses the gamma rays of higher energy that would be
reconstructed in it.

Figure~\ref{fig:sens} also shows the SVT with a 6\,m dish instead of 5\,m, with the
same camera and photodetectors (Paper~2). Its mirror area is 1.44 times larger. Its
energy threshold is 0.13\,TeV instead of 0.18\,TeV, and it is 1.3--1.7 times more
sensitive than the 5\,m dish between 0.1 and 0.63\,TeV. Above 1\,TeV, the two dishes
differ by less than 10\,\%. Between 0.1 and 0.63\,TeV, the 6\,m dish is 2.1--2.8 times
less sensitive than LST-1, and 2.0--4.5 times less sensitive than MAGIC.

The integral sensitivity in the right panel is computed from the same rates of gamma
rays and protons after all cuts. Each energy threshold is the lower edge of a bin of
estimated energy. The rates above the threshold are the sums of the rates in that
bin and in all bins above it, which requires no assumption about how the events are
distributed within a bin. The gamma-ray rates are those of a source with the
spectrum of the Crab Nebula \citep{aleksic2016}. The integral sensitivity is the
fraction of these rates, and therefore of the Crab flux above the threshold, for
which the excess meets the three criteria of Sect.~\ref{sec:sensitivity} in one
hour. It is thus directly in Crab units (C.U.). The integral of the transformer
analysis starts at 0.1\,TeV, close to its energy thresholds of 0.13 and 0.18\,TeV.
That of the standard analysis starts at 0.4\,TeV, the bin just below its threshold of
0.73\,TeV. MAGIC and LST-1 also start at 0.1\,TeV. For every curve, the bins below
these energies contain few gamma rays and much background, and including them would
make the integral sensitivity worse.

With the transformer analysis, the integral sensitivity in one hour is 37\,\%~C.U.\ above
0.16\,TeV, 29\,\%~C.U.\ above 0.25\,TeV, 25\,\%~C.U.\ above 0.4\,TeV and 22\,\%~C.U.\
above 0.63\,TeV, its best value. Above 0.25\,TeV, it is therefore 1.9 times the
indicative $\sens{15}$ of Sect.~\ref{sec:sensitivity} (dash-dotted line in
Fig.~\ref{fig:sens}, right). With photodetectors of higher QE than the photomultipliers of LST-1, the SVT
would come closer to this value: with a QE of 50\,\% (Paper~2), its integral
sensitivity above 0.25\,TeV is 20\,\%~C.U. A larger dish has a similar effect. With a
6\,m dish, the integral sensitivity is 22\,\%~C.U.\ above 0.25\,TeV, and 19\,\%~C.U.\
above 0.4\,TeV, its best value. The two improvements are independent, and their gains
should largely combine. From the gains found for each of them, we estimate that a 6\,m
station with photodetectors of 50\,\% QE would reach about 16\,\%~C.U.\ above
0.25\,TeV, close to the indicative value. The QE of 50\,\% is, however, a
simplification, and to some extent an idealisation: it is the same at all wavelengths.
Real photodetectors do not have this property. Silicon photomultipliers, for example,
have a lower efficiency below 350\,nm.

With the standard analysis, the integral
sensitivity is 250\,\%~C.U.\ above 0.4\,TeV, and its best value is 55\,\%~C.U.\ above
1\,TeV. Above 0.4\,TeV, the transformer analysis thus improves the integral sensitivity
by a factor of about ten. With the criteria of Sect.~\ref{sec:sensitivity}, the SVT with
the transformer analysis detects the Crab Nebula above 0.25\,TeV in about six minutes.
With the standard analysis, it cannot detect the Crab Nebula above 0.4\,TeV in any
observing time: the signal is only 3\,\% of the background, below the 5\,\% required. It
detects the Crab Nebula only above 0.63\,TeV, in 23 minutes.

The comparison with
the indicative $\sens{15}$ also depends on how the sensitivity is defined. The value
of about 5\,\%~C.U.\ in one hour for the current IACT systems (Sect.~\ref{sec:sensitivity}) scales their sensitivity in
long observations to one hour with $\sqrt{T}$. For MAGIC, the simpler criterion
$N_{\rm ex}/\sqrt{N_{\rm bkg}}=5$ gives 4.9\,\%~C.U.\ above 0.25\,TeV in one hour,
close to this value, whereas the definition of Fig.~\ref{fig:sens} gives 7.4\,\%. In
one hour only about seven background events remain above 0.25\,TeV, and the
criterion of Li \& Ma with five OFF regions then needs 21 excess events instead of
14. For the SVT with the transformer analysis, about 900 background events remain,
and the two definitions differ less: the criterion of Li \& Ma gives 29\,\%~C.U., and
$N_{\rm ex}/\sqrt{N_{\rm bkg}}=5$ gives 26\,\%~C.U. Compared with MAGIC above
0.25\,TeV, the SVT with the transformer analysis is 3.9 times less sensitive with the
definition of Fig.~\ref{fig:sens}, and 5.3 times with
$N_{\rm ex}/\sqrt{N_{\rm bkg}}=5$.

In summary, a stand-alone 5\,m telescope with
the photomultipliers of LST-1 comes, in simulation, within a factor of two of the
sensitivity that we found sufficient in Sect.~\ref{sec:sensitivity}, and a 6\,m dish within a
factor of 1.5. Even at the 29\,\%~C.U.\ of the present simulation, a station detects a
source of 0.2\,C.U.\ in about two hours (Eq.~\ref{eq:tdet}). For a source on the
celestial equator, the fourteen stations together then reach 4.7\,\%~C.U.\ in one day,
and 3.3--3.5\,\%~C.U.\ with 20--22\,\%~C.U.\ per station. A current IACT system on La
Palma reaches 3.0\,\%~C.U.\ in the 2.8\,h per day during which it can observe such a
source (Sect.~\ref{sec:sensitivity}).

\begin{figure*}[t]
\centering
\includegraphics[width=\textwidth]{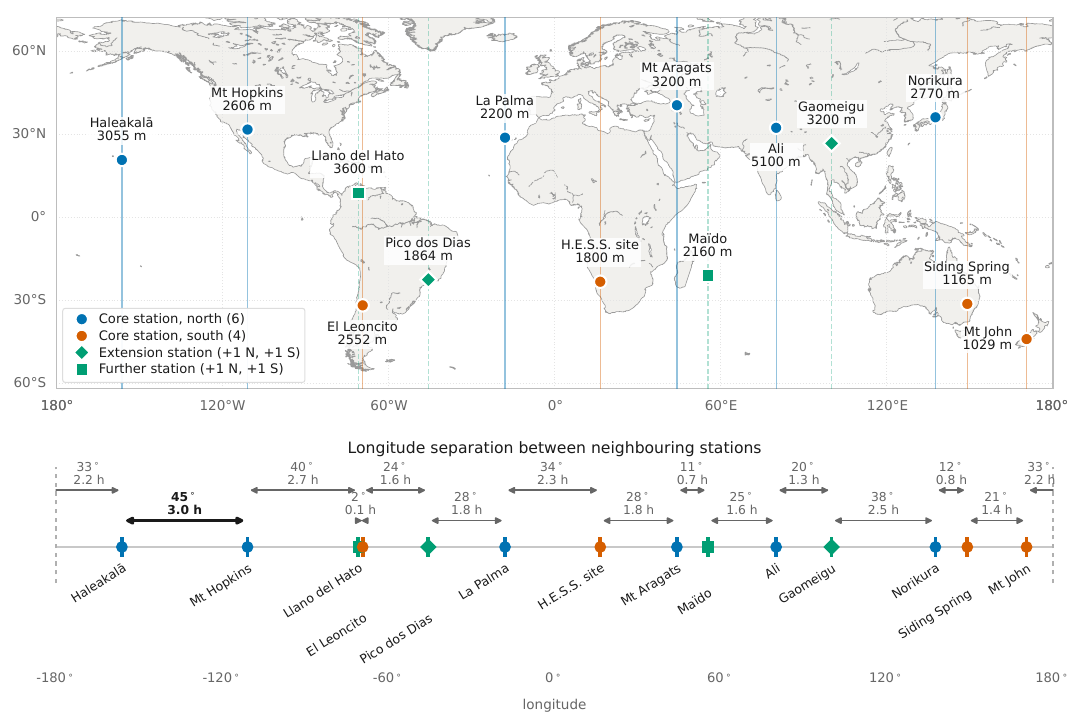}
\caption{The RevolIACT array in its fourteen-station form. All fourteen stations
are at existing observatories. Top: circles mark the
ten core stations, blue in the northern and vermilion in the southern
hemisphere; green diamonds mark the two extension stations and green squares
the two further stations; labels give the site and its altitude. Bottom:
longitude separation between neighbouring stations and the time the Earth takes
to rotate through it; the largest gap is drawn in bold. The gap across the date line
(dashed vertical lines) is split between the two edges of the panel and labelled on
both.}
\label{fig:array}
\end{figure*}

These results are the outcome of an initial study, not of a final design. The SVT
still has a perfectly reflecting and aligned mirror, a 1\,ns photodetector pulse
without transit-time spread, no electronic noise, and a background of protons only.
These protons arrive only within $4.25^{\circ}$ of the pointing direction, and their
shower axes pass only up to 250\,m from the telescope. The protons that survive the
cuts come from all distances up to this limit, so the simulated background is
probably a few tens of per cent too low. The sensitivity is affected less. Where the
background dominates, the weakest detectable flux grows only as the square root of
the background. A background that is 20\,\% too low, for example, makes this flux only
about 10\,\% too low. The low background affects both analyses, not only the
transformer. Each of these idealisations makes the simulated telescope better
than a real one.
Paper~2 describes them, and a future paper will measure how much each of them changes
the sensitivity. Other factors act in the opposite direction. Papers~1 and 2 are the first applications of a
transformer analysis to a compact telescope. Already at this early stage, the transformer analysis
lowers the energy threshold by a factor of four compared with the standard analysis,
and it is 3--17 times more sensitive between 0.4 and 1\,TeV. It will improve further,
as we learn to optimise it and as more powerful methods become available. The telescope parameters come from a preliminary scan made with an early
version of the analysis. The photomultipliers of LST-1 were
produced in 2015, and photodetectors keep improving. We are therefore confident that,
with better photodetectors and a better analysis, a station with a 5\,m diameter
mirror will reach the indicative $\sens{15}$ in the coming years. If needed, the dish can also be made
slightly larger: a 6\,m dish already comes close to this value. Two further
considerations will shape the final design.

\paragraph{The best telescope for a transformer need not be the best
telescope for the standard analysis.} The design intuition of the field was built
around an analysis that reduces each event to a few image parameters, after
discarding the faint pixels and the waveforms \citep{hillas1985}. This intuition
concerns the pixel size, the field of view, the sampling rate, and a dish large
enough to give images that the analysis can use. The transformer analysis uses
the full space-time structure of the signal, and for a given mirror area its
performance depends far less on the photodetector efficiency than that of the standard
analysis (Paper~2). The trade-offs therefore shift, and not necessarily towards
more expensive hardware. With the transformer analysis, a photodetector pulse three
times wider costs little sensitivity. Sampling at 2\,GS/s instead of 5\,GS/s,
however, raises the energy threshold from 0.18 to 0.26\,TeV, and a field of view of
$4^{\circ}$ instead of $6^{\circ}$ makes the telescope 1.2--1.8 times less sensitive
between 0.1 and 1.6\,TeV (Paper~2). Paper~2 varies these parameters only with the
transformer analysis. A future study will find out whether, and by how much, the best telescope
for a transformer analysis differs from the best telescope for the standard
analysis (Paper~2, Sect.~6).

\paragraph{Cost is a figure of merit, not an afterthought.} An array has ten or
more stations, so the quantity to optimise is the performance per
unit cost. The dominant cost is the camera, which scales with the number of
channels. For a square camera with a side of $\theta_{\rm FoV}$ and pixels of
$\theta_{\rm pix}$, this number is $N_{\rm pix}=(\theta_{\rm FoV}/\theta_{\rm pix})^{2}$,
so that halving the pixel size at a fixed field of view quadruples it. A configuration that improves the sensitivity but doubles the channel
count may be unaffordable at the scale of an array. A configuration that gives up
a small fraction of the sensitivity for a large reduction in cost may instead pay for the
stations that provide the coverage.

A dedicated design study, which optimises the sensitivity, the energy threshold and
the cost together and builds its simulations on the laboratory characterisation of
candidate components, is the subject of future work. In the rest of this paper we
assume that each station reaches the indicative sensitivity of
Sect.~\ref{sec:sensitivity}, about $\sens{15}$.

\section{The RevolIACT array}
\label{sec:array}

\subsection{Concept}

RevolIACT is an array of compact robotic stations, nominally of 5\,m, of the type
described in Paper~2. Each station is protected by an enclosure, and the stations
are distributed in longitude across both hemispheres. Its name stands for REal-sky Video-\hspace{0pt}vision-\hspace{0pt}transformer-\hspace{0pt}Optimized
Longitudinally distributed Imaging Atmospheric Cherenkov Telescopes. Each station operates
stand-alone: there is no array trigger, and no requirement that two stations record the
same shower. Besides its lower cost (Sect.~\ref{sec:compact}), a compact telescope has
four advantages over a large one. First, it can be
sheltered by a retractable enclosure, which protects the optics and stabilises the
response on which a simulation-trained analysis depends
(Sect.~\ref{sec:enclosure}). Second, it can be built to repoint anywhere on the
sky about twice as fast as large telescopes such as MAGIC and those of the CTAO LST
sub-array. This matters for GRBs,
whose VHE flux decays steeply after the trigger. Third, it can operate robotically
on a few kilowatts, which is compatible with solar power at remote sites. Fourth,
it can carry a camera with a large field of view at relatively low cost. Such a
camera stays small and light, so it does not require a heavier structure. Large telescopes have short focal ratios, for example
1.2 for LST-1 \citep{hayashida2015} and about 1 for MAGIC \citep{aleksic2016hw}. This keeps their cameras
small and light, but it degrades the images far from the centre of the camera.
The 5\,m station has a focal ratio of 1.5 (Sect.~\ref{sec:station}).
The benchmark telescope of Paper~1, which is also the
baseline of Paper~2, has a square field of view $6.1^{\circ}$ on a side, 37\,deg$^{2}$, compared with
$\sim$10\,deg$^{2}$ for MAGIC and $\sim$15\,deg$^{2}$ for LST-1
\citep{aleksic2016,abe2023}. A larger part of the error region of a
GBM or neutrino alert, often several degrees across
\citep{connaughton2015,abbasi2023icecat}, is therefore searched in a single pointing, and the much
larger regions of gravitational-wave alerts in fewer pointings.

\subsection{Layout}
\label{sec:layout}

Figure~\ref{fig:array} shows the proposed layout. It is built in three stages, of ten, twelve
and fourteen stations, which show how the coverage increases with the number of stations
(Sects.~\ref{sec:coverage} and \ref{sec:grbresults}). Ten core stations, six in the
northern and four in the southern hemisphere, define the baseline. Two extension stations,
at Pico dos Dias in Brazil and Gaomeigu in Yunnan, close the two widest longitude
gaps of that baseline, over the South Atlantic and over China between Tibet and
Japan. Two further stations, at Llano del Hato in Venezuela and Ma\"{\i}do on
R\'eunion, close the widest gap in the chain of each hemisphere. All fourteen sites
already host an observatory. Twelve lie above 1500\,m, ten above 2000\,m, and the
lowest, Mt John in New Zealand, at 1029\,m. With the ten core stations, the largest
longitude separation between neighbouring stations is $57.5^{\circ}$, which
corresponds to 3.8\,h of Earth rotation. With twelve or fourteen stations, it falls
to $45.4^{\circ}$, or 3.0\,h, and it then lies between Hawaii and Arizona. The
coordinates of all sites are listed in Table~\ref{tab:sites}. We stop at fourteen
stations because, from the \GbmBuildSmallFrom{} station on, each added station raises
the fraction of GBM bursts within reach by two percentage points or less, that is,
by at most 2\,\% of all the bursts (Sect.~\ref{sec:grbresults} and Fig.~\ref{fig:grb}). Fourteen stations are therefore
a reasonable balance between coverage and cost.

The sites are illustrative, not negotiated. They show that an array with good
geometry can be built entirely at existing observatories, not that any of them has
agreed to host a station. Appendix~\ref{app:sites} compares this layout with
fourteen stations whose locations are optimised from scratch over a list of 60
candidate sites: 45 existing observatories and 15 sites without an
observatory. The optimised layout gains less than one percentage point of
coverage and shortens the largest gap to $33$--$34^{\circ}$. An optimisation over
the existing observatories alone does almost as well, so no site without an
observatory is needed.

\subsection{Temporal coverage}
\label{sec:coverage}

\begin{figure}[t]
\centering
\includegraphics[width=\columnwidth]{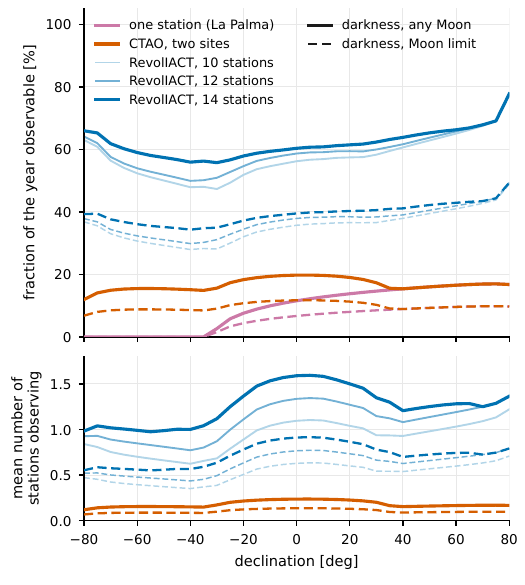}
\caption{Annual temporal coverage as a function of source declination, for La
Palma alone, the two CTAO sites and the RevolIACT array with ten, twelve and
fourteen stations (the ten- and twelve-station layouts in lighter blue). Top:
fraction of the year during which at least one station can observe the source.
Bottom: mean number of stations observing it simultaneously. Solid curves
(`darkness, any Moon') require astronomical darkness and a zenith angle below
$60^{\circ}$; dashed curves (`darkness, Moon limit') require in addition a Moon below
the horizon or less than 40\,\% illuminated. Averaged over
the sky, the fourteen stations can observe a source for 61\,\% of the year in
darkness, compared with 17\,\% for the two CTAO sites.}
\label{fig:coverage}
\end{figure}

We compute, for the year 2027 in 30-minute steps, the fraction of time during
which at least one station can observe a source, as a function of its declination
and averaged over right ascension. A station can observe when the Sun is more than
$18^{\circ}$ below its horizon and the source is at a zenith angle below
$60^{\circ}$ (Appendix~\ref{app:zenith} gives the results for limits of $30^{\circ}$
to $65^{\circ}$). In a more conservative variant, the Moon must also be below the
horizon or less than 40\,\% illuminated. H.E.S.S. observes under moonlight up to
this lunar illumination \citep{ohm2023}. MAGIC and VERITAS also observe under
brighter moonlight, at the cost of a higher energy threshold
\citep{ahnen2017moon,archambault2017}. H.E.S.S. also requires the Moon to be
$45^{\circ}$ to $145^{\circ}$ away from the source \citep{ohm2023}. We do not apply
this condition: applied whenever the Moon is above the horizon, it would lower the
coverage with the Moon limit by \MoonSepCovFourteen{} percentage points for the
fourteen stations and by \MoonSepCovCTAO{} percentage points for the two CTAO sites.
Figure~\ref{fig:coverage} shows the result for La Palma alone,
for the two CTAO sites, and for the array with ten, twelve and fourteen stations.

To average over the sky, we weight each declination $\delta$ by $\cos\delta$, so
that every part of the sky counts in proportion to its area. The coverage is then
9.5\,\% for La Palma alone and 17.3\,\% for the two CTAO sites. It rises to 56.1\,\% for the ten
core stations, 58.0\,\% for twelve and 60.8\,\% for fourteen, a factor of 3.5 over
CTAO. With the Moon limit, the values are 5.5\,\%, 10.1\,\%, 34.7\,\%, 36.6\,\% and
38.9\,\%. For a source on the celestial equator, the fourteen stations deliver on
average 14.5\,h per day of observability in darkness, compared with 4.7\,h for the two CTAO
sites and 2.8\,h for one site. The coverage also rises towards both celestial poles,
for a reason explained in Appendix~\ref{app:zenith}.

These averages hide the most useful property of the array. A source is at
opposition when it lies on the opposite side of the sky from the Sun. It is then
highest in the sky around local midnight, so that each station can observe it in
the middle of its night. Every source passes through opposition once a year, and we
say that it is near opposition within about a month of that date. Near opposition,
the stations observe the source one after another as the Earth turns. We test four
dates, the equinoxes and solstices of 2027, and five declinations, $-60^{\circ}$,
$-30^{\circ}$, $0^{\circ}$, $+30^{\circ}$ and $+60^{\circ}$, in steps of 2
minutes. In eighteen of the twenty cases, a source at opposition is observable around the clock, in
astronomical darkness, by at least one of the fourteen stations. The
two exceptions fall on the December solstice, when the nights are shortest
at the southern stations. At $-30^{\circ}$, the source is observable for 23.7
hours, with a single interruption of 18 minutes. At $-60^{\circ}$, a declination
that only the six southern stations can observe, it is observable for 21.0 hours.
Its single interruption, of 3.0 hours, follows dawn at El Leoncito. The next
stations to the west, in New Zealand and Australia, are $120^{\circ}$ and
$142^{\circ}$ away across the South Pacific, and their summer nights are short. On
the same days, the two CTAO sites cover the source
for 7 to 11 hours, depending on the date and the declination, leaving a single daily
gap of 13 to 17 hours. We repeat the test with the same twenty sources 15 and 30
days before and after opposition. The fourteen stations then cover the sources from
$-30^{\circ}$ to $+60^{\circ}$ for at least 22.9 hours per day, with interruptions
of at most 38 minutes, and those at $-60^{\circ}$ for at least 21.2 hours, with
interruptions of at most 2.8 hours. The two CTAO sites cover them for 6 to 11 hours.
The array does not only add observing hours: in most cases, it removes the daily interruption
that has limited the VHE light curves measured from any single site. With the Moon limit, this continuity is lost around
full Moon, when the Moon is above the horizon all night at every site.

Two features of Fig.~\ref{fig:coverage} should be stated plainly. First, fourteen
stations do not deliver fourteen times the coverage of one. A source must be in
the night sky at all, which for a given target is true for only part of the year,
whatever the number of sites. With the Moon limit, the stations are also limited
together around full Moon, because, at any given time, the lunar phase is the same everywhere on Earth.
Second, the mean number of stations observing the same source reaches 1.6 in
darkness and 0.9 with the Moon limit. Simultaneous observation of one source by two
stations is therefore frequent, but it is not the normal mode. It is also not
stereoscopy: stations tens of degrees apart in longitude are thousands of
kilometres apart, while a Cherenkov light pool is a few hundred metres across, so
they record different showers. When $n$ stations do observe simultaneously, the
significance grows as $\sqrt{n}$, and stations observing the same source through
different atmospheres provide a genuine cross-calibration. Simultaneous observations
also provide a contingency for campaigns coordinated with other facilities
(Sect.~\ref{sec:twosites}): a station under a clear sky can provide the sub-TeV data
when another site is covered by clouds.

\section{Prompt coverage of real gamma-ray bursts}
\label{sec:grb}

\GbmChecks The most direct test of an array built for transients is to ask how many
real transients it could have observed when they occurred.
Gamma-ray bursts are the natural test. They arrive at random times from random
directions, and they are listed in a large catalogue made with a single instrument.
Their VHE afterglow is brightest in the first minutes, when only an instrument that
already has the burst within reach can respond.

\subsection{Procedure}
\label{sec:grbmethod}

We use the \textit{Fermi} Gamma-ray Burst Monitor (GBM) Burst Catalog
\citep{meegan2009,vonkienlin2020} as retrieved from HEASARC (table FERMIGBRST) on
\GbmRetrieved. It lists \GbmN{} bursts triggered between \GbmFirst{} and
\GbmLast, \GbmPerYear{} per year on average. For each burst we take its catalogued
position and its trigger time, the time when GBM detected it. For each station we compute, at the moment of the
trigger, the zenith angle of the burst, the altitude of the Sun, and the altitude
and illuminated fraction of the Moon, using the coordinate transformations of
\texttt{astropy} \citep{astropy2022}. A burst is counted as observable by a station
if, at the trigger, it lay at a zenith angle below $60^{\circ}$ and the Sun was more
than $18^{\circ}$ below the horizon. Evaluating visibility at the trigger, rather
than over an extended window, is a good approximation. The VHE emission of a GRB is
best detected in the first ten minutes, during which the zenith angle of a burst
changes by at most $2.5^{\circ}$. We therefore also compute the visibility of each
burst 2.5, 5, 7.5 and 10 minutes after the trigger. Counting a burst as observable
if it was within reach at any of these times raises the fractions below by
\GbmTenMinLaPalma{} percentage points for La Palma, by \GbmTenMinCTAO{} for the two
CTAO sites, and by \mbox{\GbmTenMinArray} for the layouts with ten to fourteen stations.
Requiring instead that a burst stay within reach of one station at the trigger and
at the four later times
lowers them by \GbmTenMinOneLaPalma{}, \GbmTenMinOneCTAO{} and \GbmTenMinOneArray{}
percentage points, respectively. Both effects are small, so we use the visibility at
the trigger.

We compare six configurations: a single station at La Palma; the two CTAO
sites; the RevolIACT layouts with ten, twelve and fourteen stations of
Fig.~\ref{fig:array}; and fourteen optimised stations (Appendix~\ref{app:sites}).
For each configuration, we record the fraction of bursts observable by at least
one station and, for those bursts, the number of stations that could have observed
them. We also evaluate three variants. The first adds the Moon limit of
Sect.~\ref{sec:coverage}. The second requires a zenith angle below $50^{\circ}$,
because the energy threshold of an IACT rises steeply with the zenith angle
\citep[e.g.][]{aleksic2016}. Up to $50^{\circ}$, the threshold of a station with 0.075\,TeV near the zenith stays near 0.2\,TeV (Sects.~\ref{sec:sensitivity} and \ref{sec:zenith65}). The third accounts for bad weather, modelled by letting
each station observe independently on 75\,\% of occasions. This is an illustrative
figure for a good site. At the CTAO-North site on La Palma, for example, weather
prevents observations during about 20\,\% of the time \citep{gaug2024}, and 63 to
72\,\% \citep{dellavalle2010} or 80\,\% \citep{fruck2022} of the nights are clear.
In this model, a burst within
reach of $k$ stations counts with the probability $1-0.25^{k}$ that at least one of
them has a clear sky. Section~\ref{sec:zenith65} and Appendix~\ref{app:zenith} give
the results for other zenith-angle limits.

The uncertainties in Table~\ref{tab:grb} are statistical. The number of observable
bursts follows a binomial distribution, so that a fraction $f$ of the $N=\GbmN$
bursts has a standard deviation of $\sqrt{f(1-f)/N}$. In the columns of the three
variants, the number in parentheses is the gain over CTAO under the same
conditions: the fraction for the configuration divided by that for the two CTAO
sites.

\begin{table*}[t]
\caption{\textit{Fermi}-GBM bursts that could have been observed at the moment of the trigger. A burst counts as observable if at least one station had it at a zenith angle below $60^{\circ}$ during astronomical darkness (reference conditions). The three variants apply stricter or more realistic conditions; the number in parentheses is the gain over CTAO under the same conditions. The sample is the 4401 bursts of the GBM Burst Catalog triggered between 14 July 2008 and 21 October 2026 (241 per year), as retrieved from HEASARC on 23 October 2026.}
\label{tab:grb}
\centering
\small\setlength{\tabcolsep}{4pt}
\begin{tabular}{l c c c c c c c c}
\toprule
 & & \multicolumn{4}{c}{Reference: zenith angle $<60^{\circ}$, darkness} & \multicolumn{3}{c}{Variants: bursts (\%) and gain\tablefootmark{a}}\\
\cmidrule(lr){3-6}\cmidrule(lr){7-9}
 & & Bursts & Per & Gain\tablefootmark{a} & Stations & Moon & Zenith angle & Weather\tablefootmark{d}\\
Configuration & $N_{\rm st}$ & (\%) & year & & per burst\tablefootmark{b} & limit\tablefootmark{c} & $<50^{\circ}$ & \\
\midrule
La Palma alone & 1 & $9.0\pm0.4$ & 22 & 0.5 & 1.00 & 4.8 (0.5) & 6.3 (0.5) & 6.8 (0.5)\\
CTAO (North and South) & 2 & $16.6\pm0.6$ & 40 & 1 & 1.10 & 9.4 (1) & 12.5 (1) & 12.8 (1)\\
\midrule
\multicolumn{9}{l}{\textit{RevolIACT}}\\
10 core stations & 10 & $56.9\pm0.7$ & 137 & 3.4 & 1.64 & 34.5 (3.7) & 46.3 (3.7) & 48.2 (3.8)\\
12 stations (10+2) & 12 & $58.8\pm0.7$ & 142 & 3.5 & 1.91 & 36.4 (3.9) & 48.3 (3.9) & 51.4 (4.0)\\
14 stations (10+2+2) & 14 & $61.2\pm0.7$ & 147 & 3.7 & 2.15 & 38.2 (4.1) & 50.9 (4.1) & 54.5 (4.3)\\
14 stations, optimised\tablefootmark{e} & 14 & $62.3\pm0.7$ & 150 & 3.7 & 2.12 & 39.3 (4.2) & 52.1 (4.2) & 55.4 (4.3)\\
\bottomrule
\end{tabular}
\tablefoot{
\tablefoottext{a}{Fraction of bursts divided by that of the two CTAO sites under the same conditions.}
\tablefoottext{b}{Mean number of stations that could observe a burst, for the bursts observable by at least one under the reference conditions.}
\tablefoottext{c}{Reference conditions and, in addition, the Moon below the horizon or less than 40\,\% illuminated, as in Fig.~\ref{fig:coverage}.}
\tablefoottext{d}{Reference conditions, with each station independently able to observe on 75\,\% of occasions, an illustrative figure for a good site.}
\tablefoottext{e}{Layout of Fig.~\ref{fig:array14opt}, Appendix~\ref{app:sites}.}
The uncertainties are statistical (binomial, $1\sigma$; Sect.~\ref{sec:grbmethod}); for the variants, they are 0.3--0.8 percentage points.}
\end{table*}

\subsection{Results}
\label{sec:grbresults}

Table~\ref{tab:grb} gives the results, for the visibility at the trigger, without
the extension to ten minutes of Sect.~\ref{sec:grbmethod}. A single station at La
Palma could have
observed \GbmLaPalma\,\% of the bursts at the trigger, \GbmLaPalmaRate{} per year,
and the two CTAO sites together \GbmCTAO\,\%, \GbmCTAORate{} per year. Ten RevolIACT
stations raise the fraction to \GbmTen\,\%, and twelve to \GbmTwelve\,\%. Fourteen
stations reach \GbmFourteen\,\% for the layout of Fig.~\ref{fig:array} and
\GbmOpt\,\% for the optimised one: about \GbmFourteenRate{} bursts per year,
\GbmGainDark{} times as many as the CTAO sites. Almost three quarters
(\GbmNotCTAO\,\%) of the bursts within reach of the fourteen stations were out of
reach of both CTAO sites.

The gain over CTAO is robust to the observing conditions, and it grows as they
become stricter or more realistic. With the Moon limit, all the fractions fall, to
\GbmCTAOMoon\,\% for CTAO and \mbox{\GbmFourteenMoon\,\%} for the two fourteen-station
layouts, but the gain of these layouts over CTAO rises to \mbox{\GbmGainMoon}. Near full Moon, the Moon is above the horizon all
night at every site. A few days before or after full Moon, however, it sets before
dawn or rises after dusk. Every site then has a part of its night without the Moon,
and an array spread in longitude usually has a station in it. If, as at H.E.S.S.,
the Moon must also be $45^{\circ}$ to $145^{\circ}$ away from the burst whenever it
is above the horizon \citep{ohm2023}, the fraction falls by a further
\GbmMoonSepFourteen{} percentage points for the fourteen stations of
Fig.~\ref{fig:array} and by \GbmMoonSepCTAO{} percentage points for CTAO, and the gain changes only
from \GbmGainMoonFourteen{} to \GbmGainMoonSepFourteen. Restricting the zenith angle
to $50^{\circ}$ raises the gain of the two fourteen-station layouts to \mbox{\GbmGainZ}: a
burst that one station sees at a large zenith angle is often seen by another at a
smaller zenith angle. Accounting for bad weather, the gain rises to \GbmGainWeather.
A burst within reach
of the array is usually within reach of two stations at once, \GbmMult{} on
average. Of all bursts, \GbmGeTwo\,\% could have been observed by at least two
stations, which is two thirds of those within reach. Clouds at one site can
therefore often be compensated by clear sky at another, whereas the two CTAO sites
rarely share a burst (\GbmCTAOMult{} stations per burst). The fractions are
the same for short and long bursts within the statistical uncertainties. They also
agree with the sky-averaged coverage of Sect.~\ref{sec:coverage} to within about one
percentage point for every configuration. The small, systematic differences are
discussed in Sect.~\ref{sec:grbcaveats}.

\begin{figure}[t]
\centering
\includegraphics[width=\columnwidth]{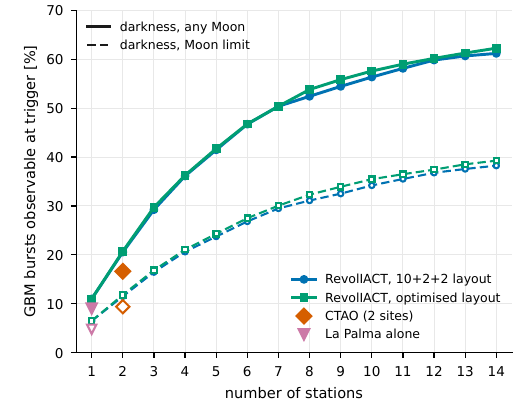}
\caption{Fraction of the \GbmN{} GBM bursts observable at the trigger as the array
grows, with stations added one at a time in the order that maximises the fraction
at each step, for the two fourteen-station layouts. Filled symbols and solid lines:
darkness, any Moon; open symbols and dashed lines: darkness, Moon limit. The diamonds show
the two CTAO sites and the triangles La Palma alone. Two stations far apart in
longitude already reach more bursts than the two CTAO sites, and six reach three
quarters of what the full array reaches.}
\label{fig:grb}
\end{figure}

Figure~\ref{fig:grb} shows how the fraction of observable bursts grows as stations are added.
The first station of the build-up, Haleakal\=a, already reaches more bursts than La
Palma: \GbmBuildOne\,\% of the bursts, compared with \GbmLaPalma\,\%. This is not
because Haleakal\=a sees more of the sky, but because GBM triggers more often over
the Pacific than over the Atlantic (Sect.~\ref{sec:grbcaveats}). Two stations far
apart in longitude can already reach more bursts than the two CTAO sites together.
Haleakal\=a and Ma\"{\i}do,
$148^{\circ}$ apart in longitude, reach \GbmBuildTwo\,\% of the bursts, compared with
\GbmCTAO\,\% for the CTAO sites, which are only $52^{\circ}$ apart. Two stations so
far apart almost never observe the same part of the sky at the same time, so that
their coverages add up: averaged over the sky, they can observe a source for
19.5\,\% of the year, the sum of what each observes alone. The two CTAO sites
sometimes observe the same part of the sky at the same time, so that their combined coverage,
17.3\,\%, is about a tenth less than the sum of their individual coverages. Both
CTAO sites also lie at the longitudes of the Americas and the Atlantic, where GBM
triggers less often. Six stations reach \GbmBuildSix\,\%, three quarters of what the full array
achieves, and from the \GbmBuildSmallFrom{} station on, each station adds two
percentage points or less. This has a practical consequence: the array can be built
in stages, and each stage is scientifically productive on its own.
Figure~\ref{fig:grb} also gives an order of construction that maximises the coverage
at each stage. In practice, the first station would be a prototype beside existing
telescopes (Sect.~\ref{sec:datasim}). A larger number
of stations also adds redundancy against bad weather or a technical problem at one
site. Without any one of the fourteen stations, the sky-averaged coverage falls by
at most 2.5 percentage points, and the fraction of GBM bursts within reach by at
most \GbmLooLoss{} percentage points. In both cases the largest loss is that of Haleakal\=a,
the only station between New Zealand and Arizona. The array therefore keeps almost
all of its coverage for the monitoring of AGN and for the hunt for transients when
one station is out of service.

The two GRBs detected by H.E.S.S., both in the GBM catalogue, show what these
fractions mean in practice. H.E.S.S. began to observe GRB\,180720B 10.1\,h after
the trigger \citep{hess2019grb}, and GRB\,190829A 4.3\,h after it \citep{hess2021grb}.
Neither burst was within reach of a CTAO site at the trigger, and La Palma could
have started to observe them only 11.4\,h and 6.5\,h after it. The fourteen
stations, instead, could have observed both bursts from the moment of the trigger,
without interruption, until H.E.S.S. began to observe them: GRB\,180720B first from
Haleakal\=a and Mt John, and GRB\,190829A first from Gaomeigu. For GRB\,190829A,
which occurred near new Moon, their coverage would even have been continuous for
the first two days, with on average two stations observing it at any time.

\subsection{Interpretation and limitations of the GBM results}
\label{sec:grbcaveats}

The fractions of Table~\ref{tab:grb} only measure whether a burst was within reach
of a station at the moment it occurred. Whether it would have been
detected at VHE depends further on its intrinsic VHE emission and on its redshift,
neither of which we model. Only a small fraction of GRBs is expected to be detectable at
VHE by any instrument, and the more sensitive CTAO will detect fainter or more
distant bursts among those it can observe. The array multiplies the number of bursts
that can be observed. For the brightest bursts, which in their first minutes are far
above the sensitivity of a single station (Table~\ref{tab:cases}), this number is
what limits the detection rate.

The GBM sample is not quite an isotropic population. GBM does not trigger while
\textit{Fermi} crosses the South Atlantic Anomaly, which removes bursts above
the Americas and the Atlantic. In the catalogue, bursts are up to
\GbmSaaDeficit\,\% less frequent than average at those longitudes, and up to
\GbmPacificExcess\,\% more frequent over the western Pacific. Stations in the
Pacific, such as Haleakal\=a, therefore reach more GBM bursts than their sky
coverage implies, and the CTAO sites fewer. Averaged over the sky, for example,
Haleakal\=a and La Palma can observe a source for almost the same fraction of the
year, 9.7\,\% and 9.5\,\%, but Haleakal\=a reaches \GbmBuildOne\,\% of the GBM bursts
and La Palma \GbmLaPalma\,\%. Together with the latitude of the site,
this spreads the single-station fractions from \GbmSingleMin{} to
\GbmSingleMax\,\% across the sites of Table~\ref{tab:sites}. For an isotropic
population, such as the counterparts of gravitational waves, the gain
of fourteen stations over CTAO is that of the sky-averaged coverage of
Sect.~\ref{sec:coverage}: \GbmIsoGainDark{} in darkness and \GbmIsoGainMoon{} with the
Moon limit, about \GbmIsoLess\,\% less than for the GBM bursts. Neutrino alerts are less uniform,
because the sensitivity of a neutrino telescope depends on the direction from which
the neutrino arrives \citep{abbasi2023icecat}. Finally, the GBM localisations of the
bursts are uncertain by
several degrees. The uncertainty shifts the zenith angle of individual bursts at
random, and so does not bias the fractions. For the follow-up itself, such an error
region is larger than the field of view of any IACT, and a wide camera reduces the
number of pointings needed to cover it.

The Einstein Telescope will announce some neutron star mergers minutes before they
occur \citep{banerjee2023}. A station that has such a merger within reach can then be
pointing at the region of the alert when the merger happens. With fourteen stations,
this is the case for 61\,\% of these mergers, compared with 17\,\% for the two CTAO
sites. The array is also more robust against bad weather, which matters most for such
singular events (Sect.~\ref{sec:singular}). If a merger is within reach of only one
CTAO site and the sky there is cloudy, the event is lost. Of the GBM bursts within reach
of the fourteen stations, instead, two thirds could have been observed by at least two
stations (Sect.~\ref{sec:grbresults}), so that a station under a clear sky is often
available. In the weather model of Sect.~\ref{sec:grbmethod}, bad weather removes
23\,\% of the bursts within reach of the CTAO sites, but only 11\,\% of those within
reach of the fourteen stations (Table~\ref{tab:grb}).

\subsection{Results for other zenith-angle limits}
\label{sec:zenith65}

Unless stated otherwise, the results above assume a zenith angle below $60^{\circ}$.
Appendix~\ref{app:zenith} repeats the main calculations for limits of
$30^{\circ}$, $40^{\circ}$, $50^{\circ}$ and $65^{\circ}$. Some observations are
made at larger zenith angles: MAGIC, for example, observes the Galactic centre from
La Palma at zenith angles of $58^{\circ}$ to $70^{\circ}$ \citep{ahnen2017}. With a
limit of $65^{\circ}$, every site sees more of the sky. For the GBM bursts, the
fraction observable at the trigger rises from \GbmCTAO{} to
\GbmCTAOSixtyFive\,\% for the two CTAO sites, and from \GbmFourteen{} to
\GbmFourteenSixtyFive\,\% for the fourteen stations. The gain of the array over
CTAO falls slightly, from \GbmGainDark{} to \GbmGainDarkSixtyFive. The extra five
degrees increase the sky coverage of one or two sites by 12--15\,\%, but that of the
array, which already covers most of the night sky, by only 8\,\%. Stricter limits
have the opposite effect. Every fraction falls, but the gain of the fourteen
stations of Fig.~\ref{fig:array} over CTAO grows, to \GbmGainFifty{} at
$50^{\circ}$, the variant of Table~\ref{tab:grb}, and \GbmGainThirty{} at
$30^{\circ}$ (Table~\ref{tab:zenith}). These
gains apply the same limit to CTAO. The large telescopes of the CTAO LST
sub-array, however, keep an energy threshold below 0.1\,TeV up to zenith angles of
$40$--$50^{\circ}$, so that CTAO suffers much less than a compact station from
observations at large zenith angles. Therefore, with strict zenith-angle limits, such
as $30^{\circ}$, these gains overstate what the array of compact telescopes adds to
CTAO.

We keep $60^{\circ}$ as the reference, because up to this angle the threshold of a
station stays within the sub-TeV band. The threshold rises steeply with zenith
angle: it is about 5 times higher at $60^{\circ}$ than near the zenith, and 7 times
higher at $65^{\circ}$ (Appendix~\ref{app:zenith}). The factor is similar for large
and small telescopes, but its consequence is not. A station with a threshold of
0.075\,TeV near the zenith would reach about 0.2\,TeV at $50^{\circ}$, 0.37\,TeV at
$60^{\circ}$ and 0.54\,TeV at $65^{\circ}$. This threshold is a goal
(Sect.~\ref{sec:sensitivity}). With the 0.13\,TeV that Paper~2 finds for a 6\,m dish with
the photomultipliers of LST-1, a station would reach 0.2\,TeV at about $35^{\circ}$,
0.36\,TeV at $50^{\circ}$ and 0.64\,TeV at $60^{\circ}$. A large telescope with a threshold of a
few tens of GeV near the zenith stays well below 1\,TeV. For a compact station, the
additional coverage between $60^{\circ}$ and $65^{\circ}$ is therefore of limited
use. For the faint transients of Table~\ref{tab:cases}, whose fluxes are quoted
above 0.2\,TeV, the useful limit is about $50^{\circ}$ for a station that reaches this
goal. Fourteen stations then still
observe a source 51\,\% of the time, three times the coverage of the two CTAO sites
with the limit of $60^{\circ}$ (Table~\ref{tab:zenith}).

\section{Feasibility}
\label{sec:feasibility}

\subsection{Idealised simulations}
\label{sec:idealised}

The performance on which this paper builds comes
from simulations that are still idealised and simple (Sect.~\ref{sec:station}). A
more detailed simulation of the station, with realistic optics, photodetector timing, electronics and cosmic-ray background,
may change its expected performance somewhat. Better photodetectors and a better
analysis act in the opposite direction (Sect.~\ref{sec:station}). The largest challenge in realising a station, however, lies elsewhere: in
the agreement between the real data and the simulations on which the analysis is
trained.

\subsection{Agreement between data and simulation}
\label{sec:datasim}

A network trained on simulations
depends on the fidelity of the simulated instrument more than the standard
analysis does, precisely because it uses more of the information in the data
(Paper~1). If the real telescope differs from the simulated one, for example in its
point-spread function, mirror reflectivity, pixel gains or night-sky background, the
network may perform worse on real data than in simulation. This mismatch, known in
machine learning as domain shift, is a central issue for every deep-learning analysis
of IACT data
\citep[e.g.][]{shilon2019,miener2024}, and it is the principal risk of the concept.
The transformer analysis has not yet been applied to real data. The present IACTs
are not well suited to this test. Their hardware is exposed to the weather, so that their optical response drifts
\citep{abe2023}. Their simulations were tuned for the standard analysis, which
does not need such an accurate model of the instrument. Their cameras have hexagonal
pixels, which would have to be resampled onto the square grid that the transformer
expects, and smaller fields of view (Sect.~\ref{sec:array}). The risk is therefore best
tested with a station designed for this purpose from the start, and it can be
reduced in six ways:
\begin{enumerate}
\item Accurate simulation and digital twin. Every component of the telescope is
  measured in the laboratory and modelled accurately in the Monte Carlo simulation.
  Together, these models form a digital twin of the whole telescope, that is, a
  detailed computer model of it, which is then used to test and evaluate the
  analysis, the calibration and the operation of the station.
\item Requirements before construction. Simulations with deliberately distorted
  instrument responses (point-spread function, mirror reflectivity, pixel gains,
  night-sky background) show how much each mismatch costs in sensitivity. They set
  the accuracy required of the calibration system. This simulation study is planned
  for a future paper.
\item Domain adaptation. The network is adapted to the real data with established
  machine-learning techniques. It can be fine-tuned, that is, trained further, on
  real events whose type is known, such as muon rings, the excess of gamma rays from
  bright sources and cosmic-ray electrons. Domain-adversarial training
  \citep{ganin2016} makes the network less sensitive to the differences between data
  and simulation. The free parameters of the simulation can be tuned, so that the
  internal representations of the network agree between data and simulation.
  Architectures that estimate their own uncertainty can flag events unlike any seen
  in training.
\item Cross-calibration. A station placed beside operating IACTs can compare its
  flux measurements with theirs. When it records the same air showers, it can also
  compare its reconstruction of individual events.
\item Stability by design. The enclosure (Sect.~\ref{sec:enclosure}) keeps the mirror reflectivity,
  the alignment of the mirrors and of the camera, and the humidity inside the camera
  stable, so that the response of the station changes little with time.
\item Atmospheric monitoring. The atmosphere, the one part of the instrument that
  cannot be enclosed, is monitored continuously, and the data are corrected for its
  transmission, as MAGIC does with its lidar \citep{schmuckermaier2023}.
\end{enumerate}
The challenge is now to demonstrate that such a station can be built, and that the
transformer analysis keeps much of its performance on real data. A prototype station, operated beside existing IACTs (item 4),
can provide this demonstration.

If the measures above do not suffice, two options remain, with some loss of
performance: a hybrid analysis that combines the transformer with image parameters,
or retraining on simulations anchored to the data.

\subsection{Enclosure}
\label{sec:enclosure}

The present IACTs are exposed to the weather all year round:
to wind, rain, dust and sand, snow, and large differences in temperature between
day and night. Their mirrors lose a few per cent of their reflectivity per year and
must be re-coated, refurbished or replaced after some years, which is expensive and
time-consuming \citep{forster2013,will2019}. Exposure to harsh weather is thus one
of the main sources of maintenance cost for the present IACTs. At some sites it
includes volcanic ash. The eruption of the Cumbre Vieja volcano on La Palma, from September
to December 2021, brought ash fall and considerable downtime to the Roque de los
Muchachos Observatory, although no installation was damaged
\citep{perezarencibia2022}. It interrupted the operation of LST-1. After the
eruption, the optical efficiency of LST-1 was about 5\,\% lower. It recovered only
partly after rain in February 2022, which presumably washed dirt off the dish
\citep{abe2023}. A retractable
hangar on rails parks the telescope under cover by day and in bad weather. The CAT
telescope, whose 17.8\,m$^{2}$ reflector is in exactly this size class, used such a
hangar \citep{barrau1998}. The hangar protects the mirrors, the structure and the
camera, and so reduces maintenance. It also keeps the optical throughput, the
alignment of the mirrors and the camera environment stable, which serves the
agreement between data and simulation. A hangar of this kind is simpler and less expensive
than a rotating dome, and it gives easy access to the mirrors and the camera. A
retractable enclosure is also planned for the TDT (Sect.~\ref{sec:intro}): each
cluster of its 2\,m telescopes would be housed in a low building with a retractable
roof \citep{kupfer2025}.

\subsection{Robotic operation, power and AI agents}
\label{sec:robotic}

A compact telescope is safe to
operate unattended and is expected to consume less than 3--4\,kW, which makes solar
power possible and remote sites affordable to run. Robotic operation also allows an
automatic response to alerts. The network is small by current standards, about
seven million parameters. Even its training processes about 160 events per second
on a single GPU (Paper~1), and applying the trained network is faster. Reconstructing the data on site,
and issuing alerts within minutes of the start of a flare, is therefore a realistic
goal for each station. Each station could be run by its own AI agent, which
schedules its observations, monitors the weather and the hardware, and reacts to
alerts. A central agent would coordinate the station agents. It would decide which
station observes which source, hand a source over from one station to the next as
the Earth turns, and redistribute the observations when a station is covered by clouds or
out of service. Such an architecture has precedents. In the eSTAR project, software
agents placed observations on a network of robotic telescopes and followed up
alerts without human intervention \citep{allan2008}. The global network of
robotic telescopes of the Las Cumbres Observatory is scheduled centrally
\citep{brown2013}. Agents based on large language models now plan the observations
of a supernova survey with ten small optical telescopes, coordinated by a central
agent \citep{wang2025}, and control an operational solar telescope
\citep{tong2026}. Similar agents have been proposed for the control software and
the data analysis of CTAO \citep{kostunin2025}. The TDT (Sect.~\ref{sec:intro}) is
planned with AI-driven, flexible scheduling, which would combine regular
observations with targets of opportunity, both urgent and less urgent, and respond
within a minute.\footnote{Summary of the TDT concept:
\url{https://www.astro.dur.ac.uk/~simo/TDTelescope/images/tdt_summary.pdf}} Its
designers consider the scheduling possibly the hardest part of the project
\citep{kupfer2025}.

\subsection{Alert triage}
\label{sec:triage}

The central agent would also decide which alerts the stations follow. The Rubin
Observatory alone is expected to issue of order $10^{7}$ alerts per night
\citep{ivezic2019}. Radio surveys, X-ray and gamma-ray satellites, and neutrino and
gravitational-wave observatories add their own (Sect.~\ref{sec:transients} and
Table~\ref{tab:alerts}). The
Einstein Telescope alone is expected to detect more than $10^{5}$ binary mergers per
year, each of them a potential alert \citep{maggiore2020}. Nobody can inspect these
streams by eye. Alert brokers are being developed to process them. They receive the
alerts, cross-match them with other catalogues, classify them, often with machine
learning, and select those worth observing further. The Rubin Observatory has seven
community brokers \citep{vujcic2025}, among them ALeRCE \citep{forster2021}, AMPEL
\citep{nordin2019}, ANTARES \citep{matheson2021} and Fink \citep{moller2021}.
AMPEL is developed close to the gamma-ray and neutrino communities, at the Humboldt
University of Berlin and at DESY in Zeuthen, where the Science Data Management
Centre of CTAO is located. It is used in particular to search the alerts of ZTF for
the optical counterparts of high-energy neutrinos \citep{stein2023} and of
gravitational-wave events \citep{kasliwal2020}. The central agent would take the
alerts selected by such brokers, rank them by their scientific interest and by the
chance of a VHE detection, and assign the most promising ones to the stations that
can observe them, without waiting for a human decision. This ranking is a natural
extension of the machine-learning programme and is left to future work.

\subsection{Siting}
\label{sec:siting}

Distribution in longitude means international siting, with the
host agreements, logistics and politics that this implies. The layout of
Fig.~\ref{fig:array} uses only existing observatories, precisely so that roads,
power, network and local support already exist, and Appendix~\ref{app:sites} shows
that this restriction costs almost nothing in coverage. In practice, an array of
this kind would be built by an international collaboration, station by station, and
Fig.~\ref{fig:grb} shows that each stage adds science on its own.

\section{What stand-alone operation gives up}
\label{sec:limits}

\subsection{Sensitivity per station}
\label{sec:sensperstation}

A station at the indicative sensitivity of Sect.~\ref{sec:sensitivity} is about three
times less sensitive than the current major IACT systems, and the gap to CTAO is
even larger. The array is
complementary to CTAO, not a competitor. CTAO provides deep observations where it
can point. The array provides observations where and when CTAO cannot. An alert
detected by a station can in turn trigger deeper observations by CTAO when the
source becomes observable from a CTAO site.

\subsection{Stereoscopy}
\label{sec:stereo}

When several telescopes observe the same shower, the
combination of their images reconstructs the event better than a single image,
especially with three or four telescopes. Besides a better sensitivity, this yields
better energy and angular resolutions. A stand-alone station sees each shower
from one direction only. The time structure of
the waveforms carries part of the information that stereoscopy provides, notably on
the impact distance of the shower, and an analysis of the waveforms can use it. Even so, stereo arrays keep these advantages. Apart from its contribution to the sensitivity, a good
angular resolution matters most for the morphology of extended sources, such as
pulsar wind nebulae and supernova remnants, which CTAO will map in great detail.
RevolIACT, instead, is designed for time-domain astronomy. Its targets are
point-like sources, in most cases detected within minutes to hours, so it gains
little from a good angular resolution, unlike CTAO, for which much of the science
concerns extended sources.

\section{Conclusions}
\label{sec:conclusions}

The sub-TeV band is about to become more sensitive. At the same time, it may become
less able to watch the sky continuously. By the mid-2030s, IACTs with good
sensitivity down to $\sim$0.1\,TeV may operate at only two locations, the two CTAO
sites. This happens
just as the number of transients worth watching grows by orders of magnitude. We
have quantified how much of this gap could be closed by RevolIACT, an array of compact,
stand-alone Cherenkov telescopes distributed in longitude, inexpensive enough to be
built in numbers and made scientifically useful by the transformer analysis.

An integral sensitivity per station of order $\sens{15}$ above 0.2--0.3\,TeV,
about three times worse than that of the current IACT systems, would be sufficient. With it, the coverage of the stations turns into detections of the transients that
matter. A source of 0.1\,C.U.\ above 0.2\,TeV, such as the weak but nearby
GRB\,190829A on its first night, is detected within a night. Fainter sources, such as
the neutrino counterpart TXS\,0506+056, are detected within a day by combining the
stations. A sensitivity
about nine times worse would not detect the fainter ones in time. Useful sensitivity down
to $\sim$0.1\,TeV adds more distant sources. With the standard analysis, a
single 5\,m telescope with the photomultipliers of LST-1 reaches only 250\,\%~C.U.\
above 0.4\,TeV in one hour, 17 times the indicative $\sens{15}$. With the transformer
analysis, in a still idealised simulation, the same telescope
reaches 29\,\%~C.U.\ above 0.25\,TeV in one hour, within a factor of two of the indicative value. It reaches 20\,\%~C.U.\ with
photodetectors of 50\,\% QE, and 22\,\%~C.U.\ with a 6\,m dish
(Sect.~\ref{sec:station}). Together, the two improvements should bring a 6\,m station
to about 16\,\%~C.U. The QE of 50\,\% is, however, a simplification: it is the same at
all wavelengths. Papers~1 and 2 are the first applications of a transformer analysis to
a compact telescope. Compared with the standard analysis, the transformer analysis already lowers the
energy threshold by a factor of four, and it is more sensitive at all energies. Both the analysis and the photodetectors will
improve. We are therefore confident that a station with a 5\,m diameter mirror will reach the
indicative $\sens{15}$ in the coming years. A slightly larger dish would also help: a 6\,m dish
already comes close to this value. The design of the station has not yet been
optimised. A future study will jointly optimise the sensitivity, the energy
threshold and the cost.

With ten to fourteen stations at existing observatories, a source can be observed
56--61\,\% of the time, averaged over the sky and the year, compared with 17\,\% for
the two CTAO sites. With fourteen stations, most sources near opposition, within about
a month of the date on which they are opposite the Sun, can be followed around the
clock. Far in the southern sky, around the December solstice, a daily gap of up to
three hours remains.
Without any one station, the array keeps almost all of this coverage. Tested on
\GbmYears{} years of real gamma-ray bursts, the fourteen stations of
Fig.~\ref{fig:array} could have observed \GbmFourteenIntOne\,\% of the \GbmN{} GBM
bursts at the trigger, compared with \GbmCTAOInt\,\% for the CTAO sites:
\GbmGainDark{} times as many. The gain rises to \GbmGainMoonFourteen{} when
observations also require the Moon to be below the horizon or less than 40\,\%
illuminated, and to \GbmGainWeatherFourteen{} when accounting for bad weather. For sources spread uniformly
over the sky, such as the counterparts of gravitational waves, the
gain of fourteen stations is that of the sky-averaged coverage: \GbmIsoGainDark{} in
darkness and \GbmIsoGainMoon{} with this Moon limit. This is about \GbmIsoLess\,\%
less than for the GBM bursts, because GBM triggers less often at the longitudes of
the Americas and the Atlantic, where both CTAO sites lie, and more often over the
Pacific. Six well-placed stations already reach three quarters of the
bursts that the full array reaches. The two GRBs detected by H.E.S.S., hours after
the burst, could have been observed by the stations from the moment of the trigger,
and the afterglow of GRB\,190829A could have been detected continuously until about
one day after the burst. RevolIACT would complement CTAO, adding
temporal coverage to its sensitivity (Sect.~\ref{sec:sensperstation}).

The decisive next step is empirical: to show that the transformer analysis keeps
its large advantage on real data. The transformer is trained on simulated
events, so it performs as well on real data only if the simulation reproduces the
real telescope accurately. The telescope must therefore be designed for this purpose
from the start.
An enclosure keeps its mirrors, their alignment and its camera stable, so that its
response changes little with time. Laboratory measurements of every component,
combined in a digital twin of the telescope, and continuous calibration allow this
response to be simulated in detail
(Sect.~\ref{sec:datasim}). This accuracy was not a design goal of the present
IACTs, whose standard analysis does not need it. A single prototype station of this
kind, operated beside existing telescopes that observe the same sources and provide
reference measurements, can provide this test.

\begin{acknowledgements}
\textit{Author contributions.} D.~Paneque: conceptualisation of RevolIACT;
supervision; computation of the temporal coverage of the array layouts under
different observing conditions, including the test on the GBM bursts;
visualisation of the results; project administration; funding acquisition;
writing -- original draft, review and editing. E.~Jobst: main developer of the
transformer analysis; Monte Carlo simulations of the telescope configurations;
training and validation of the transformer analysis and of the standard analysis
for these configurations; writing -- review and editing. L.~Heinrich: methodology
and supervision of the transformer analysis; funding acquisition; writing -- review
and editing. A.~Hahn: conceptualisation, through many discussions of the RevolIACT
concept; a detailed review that shaped the structure and the content of the paper;
writing -- review and editing.

We are very grateful to R.~Mirzoyan, who provided a detailed and constructive
list of remarks at short notice, on the last day before submission. We also
thank A.~Berti, A.~Moralejo and R.~L\'opez-Coto for discussions and for
information on the sensitivity of LST-1. We are grateful to the developers of
the CTAO \texttt{simtools} package, a reliable and easy-to-use code that we used
extensively for our simulations. The computations were performed on the
high-performance computing clusters of the Max Planck Computing and Data Facility
(MPCDF) and of the Origins Data Science Lab (ODSL). D.P. acknowledges support from
the Deutsche Forschungsgemeinschaft (DFG, German Research Foundation) under Germany's
Excellence Strategy -- EXC-2094 -- 390783311. E.J. and L.H. are supported by the BMFTR project
SciFM (05D25WO2). This work made use of the \textit{Fermi} GBM Burst Catalog
provided by the High Energy Astrophysics Science Archive Research Center (HEASARC)
at NASA Goddard Space Flight Center, and of \texttt{astropy} \citep{astropy2022}.
\end{acknowledgements}

\bibliographystyle{aa}
\bibliography{references}

\begin{appendix}

\section{Choice of the station sites}
\label{app:sites}

\subsection{Method}

The layout of Fig.~\ref{fig:array} was built incrementally: ten core stations at
existing observatories, then two stations that close the widest gaps of that
baseline, then two more that close the widest gap in the chain of each hemisphere.
To see how far this is from the best that fourteen stations can do, we also
optimise the locations of the fourteen stations from scratch.

The candidates are 60 sites, compiled by hand to cover all longitudes in
both hemispheres. Forty-five of them are existing observatories. Forty-three of
these are astronomical or astroparticle observatories, and two, at Ma\"{\i}do on
R\'eunion and at Pico Espejo in Venezuela, are atmospheric research stations. The
other fifteen are remote islands or mountains without an observatory, included
because they fill longitudes where no observatory exists. For each of them, we take
the coordinates and altitude of a named point, in most cases the summit. We check
the coordinates and altitudes of all 60 candidates against published values.
Column~5 of Table~\ref{tab:sites} marks the existing observatories.

The optimisation looks for the set of fourteen candidates with the best figure of
merit, $J$, which has two parts. The first is the coverage $C$: the fraction of the
year during which at least one station can observe a source, with the Moon limit,
averaged over the sky as in Sect.~\ref{sec:coverage}. The second is a penalty for a
wide gap in the chain of stations. The coverage is an average over the sky and
the year, and it changes little when one gap between neighbouring stations becomes
wide. Such a gap, however, interrupts the light curve of every source once a day
(Sect.~\ref{sec:coverage}). A fast transient that occurs during this interruption,
for example the prompt emission of a GRB, is lost, because it is over before any
station can observe it. The largest longitude gap, $\Delta\lambda_{\max}$, cannot
be smaller than $360^{\circ}/14=25.7^{\circ}$, the gap between fourteen equally
spaced stations, and the penalty is proportional to the excess over this value:
\begin{equation}
J \;=\; C-0.36\,\frac{\max(0,\,\Delta\lambda_{\max}-25.7^{\circ})}{360^{\circ}}.
\label{eq:objective}
\end{equation}
The weight 0.36 is a choice, not the result of a fit. With it, each degree by which
the largest gap exceeds $25.7^{\circ}$ costs as much as 0.1 percentage point of
coverage, so that closing a gap by $10^{\circ}$ is worth one percentage point.

\begin{figure*}[t]
\centering
\includegraphics[width=\textwidth]{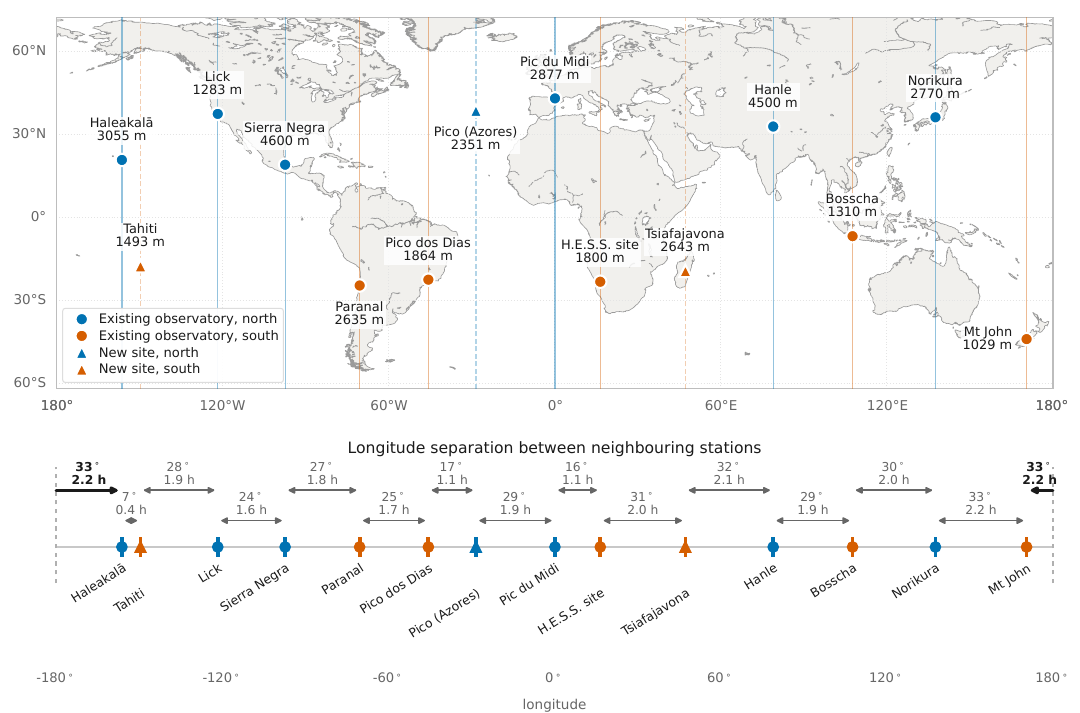}
\caption{Fourteen stations with locations optimised from scratch over all
candidate sites at 900\,m or higher. Circles mark existing observatories
and triangles sites without an observatory; blue marks the northern
and vermilion the southern hemisphere. Bottom panel as in Fig.~\ref{fig:array}.}
\label{fig:array14opt}
\end{figure*}

The search proceeds in two steps. First, we add stations one at a time, each time
the candidate that raises $J$ most, until fourteen are chosen. Second, we try
replacing each chosen site by each unused candidate, and keep any replacement that
raises $J$, until none does.
Such a search finds a very good layout, but not necessarily the best of all
possible ones. To keep it fast, it used a coarse grid (hourly steps through 2027,
declinations every $10^{\circ}$, twelve right ascensions), a zenith-angle limit of
$60^{\circ}$ and a minimum altitude of 900\,m, which 47 of the candidates meet, 42 of
them existing observatories. We ran the optimisation over all 47 and over the 42
existing observatories only. Section~A.3 adds two runs over all candidates, without
a minimum altitude and with a minimum altitude of 1500\,m. We call the result of each
run an optimised layout. Figure~\ref{fig:array14opt} shows the optimised layout over
all candidates with the minimum altitude of 900\,m. Table~\ref{tab:layouts} compares the optimised
layouts with the incremental layout on the fine grid of Sect.~\ref{sec:coverage},
and Table~\ref{tab:sites} lists all 60 candidates.

\begin{table*}[t]
\caption{The fourteen-station layouts compared. The incremental layout is that of Fig.~\ref{fig:array}; the optimised layouts result from the optimisation of Appendix~\ref{app:sites} with different constraints. Coverage is the fraction of the year during which at least one station can observe a source, averaged over right ascension and over declination weighted by $\cos\delta$; GBM is the fraction of the bursts of Table~\ref{tab:grb} observable at the trigger. Zenith angle below $60^{\circ}$ throughout.}
\label{tab:layouts}
\centering
\small\setlength{\tabcolsep}{5pt}
\begin{tabular}{l c c c c c}
\toprule
 & Incremental & \multicolumn{4}{c}{Optimised} \\
\cmidrule(lr){3-6}
 & (10+2+2) & Existing observatories & Sites without & No minimum & Altitude \\
 & & only, $\geq$900\,m & observatory allowed, $\geq$900\,m & altitude & $\geq$1500\,m \\
\midrule
Coverage, darkness (\%) & 60.8 & 61.3 & 61.5 & 62.8 & 56.2\\
Coverage, Moon limit (\%) & 38.9 & 39.4 & 39.5 & 40.6 & 36.4\\
GBM bursts, darkness (\%) & 61.2 & 61.5 & 62.3 & 63.3 & 56.4\\
GBM bursts, Moon limit (\%) & 38.2 & 38.8 & 39.3 & 40.2 & 35.9\\
Largest longitude gap & $45^{\circ}$ (3.0\,h) & $34^{\circ}$ (2.3\,h) & $33^{\circ}$ (2.2\,h) & $39^{\circ}$ (2.6\,h) & $66^{\circ}$ (4.4\,h)\\
Sites below 900\,m / 1500\,m & 0 / 2 & 0 / 4 & 0 / 4 & 4 / 5 & 0 / 0\\
Sites without an observatory & 0 & 0 & 3 & 3 & 1\\
\bottomrule
\end{tabular}
\end{table*}

\subsection{Advantages and disadvantages}

Optimising from scratch buys continuity rather than coverage. The largest
longitude gap falls from $45^{\circ}$ (3.0\,h of Earth rotation) to
$33$--$34^{\circ}$ (2.2--2.3\,h), while the coverage rises by only 0.5--0.7
percentage points and the fraction of GBM bursts by \mbox{\GbmOptDiff}
(Table~\ref{tab:layouts}). Most of the gain comes from one region. The incremental
layout has a single station, Mt Hopkins, between Hawaii and South America, and its
two widest gaps, $45^{\circ}$ and $40^{\circ}$, lie on either side of it. Both
optimised layouts with the minimum altitude of 900\,m place two stations there, at Lick in
California and Sierra Negra in Mexico. Their largest gap is then the central Pacific
between New Zealand and Hawaii. Both are also balanced between the hemispheres, with
seven stations in each, compared with eight and six. At opposition, each optimised layout
keeps a source in view around the clock in nineteen of the twenty cases of
Sect.~\ref{sec:coverage}. The layout restricted to existing observatories falls to
21.6 hours at $-60^{\circ}$ on the December solstice, compared with 21.0 hours for the
incremental layout. The one over all candidates falls to 23.9 hours at $+60^{\circ}$
on the June solstice.

The optimised layout over all candidates includes three sites without an
observatory, Mt Marau on Tahiti, Pico in the Azores and Tsiafajavona in Madagascar,
but they raise the coverage by only 0.2 percentage points. The optimised layout restricted to existing
observatories reaches 61.3\,\% coverage with a largest gap of $34^{\circ}$, compared
with 61.5\,\% and $33^{\circ}$ when such sites are allowed, and \GbmOptExist\,\% of
the GBM bursts, compared with \GbmOpt\,\%. It shares nine of its fourteen sites with
the layout over all candidates, and replaces the other five by Mauna Kea, Teide,
Skinakas, Ma\"{\i}do and Siding Spring.

The practical disadvantages of the optimised layouts are of a different kind. They
sit lower:
four of their sites lie below 1500\,m, compared with two in the incremental layout, and
seven or eight above 2000\,m, compared with ten. Finally, the optimisation does not take
into account the weather at the sites or the artificial light around them. Lick, for instance, overlooks the San Francisco Bay
Area. Replacing it in the optimised layout restricted to existing observatories by
San Pedro M\'artir in Baja California lowers the coverage by only about 0.1 percentage point, but widens the
largest gap from $34^{\circ}$ to $40^{\circ}$. Coverage is insensitive to the choice
of individual sites, but the worst-case gap is not.

Three conclusions follow. First, the sites can be chosen on practical grounds, such
as host agreements, access, sky quality and weather, at almost no cost in coverage.
The incremental layout and the optimised layout restricted to existing observatories
share fewer than half of their sites, but differ in coverage by only half a
percentage point. Second, the one geometric
lesson worth keeping is to place two stations, not one, between Hawaii and South
America. Third, no site without an observatory is needed. The
incremental layout and the optimised layout restricted to existing observatories
both include a station in the Canary Islands, at La Palma and at Teide respectively,
where a prototype can be operated beside existing telescopes
(Sect.~\ref{sec:datasim}).

\subsection{The altitude constraint}

The minimum altitude matters more than the choice of individual sites
(Table~\ref{tab:layouts}). Without a minimum altitude, the optimised layout reaches the
highest coverage, 62.8\,\% in darkness and \GbmOptNone\,\% of the GBM bursts, by placing four
stations at about 500\,m or lower: on Wake Island, on the Chatham Islands, on Easter
Island and at Murchison. The altitude of an IACT affects its energy threshold, so that this
trade gives up precisely the sub-TeV performance on which the concept depends. Its
largest gap, $39^{\circ}$, is also worse than with a minimum altitude of 900\,m. With a
minimum altitude of 1500\,m, the sites of Australia and New Zealand, all below that altitude, drop out.
The largest gap then grows to $66^{\circ}$, between Japan and Hawaii, and the
coverage falls to 56.2\,\%, below that of the incremental layout. A minimum altitude
of about 900--1000\,m is a reasonable compromise, and it is the one adopted here.

\subsection{A larger penalty for the largest gap}
\label{app:penalty}

The weight 0.36 in Eq.~(\ref{eq:objective}) is a choice. We therefore repeat the
four optimisations with a weight five times larger, 1.8, so that each degree by
which the largest gap exceeds $25.7^{\circ}$ costs as much as 0.5 percentage point of
coverage. The search of Sect.~A.1 is not exhaustive. Its second step replaces one
site at a time, and it stops when no single replacement raises $J$. A better layout
may then still exist, but it can be reached only by replacing several sites at
once. With the weight of 1.8, the search of Sect.~A.1 over all candidates at 900\,m
or higher stops at such a layout, with a largest gap of $34^{\circ}$, wider than the
$33^{\circ}$ found with the weight of 0.36. We therefore
also start the second step from the four layouts of Table~\ref{tab:layouts} and
from four random sets of fourteen sites, and keep the layout with the highest $J$.
With the minimum altitude of 900\,m, the larger weight then changes the result only slightly. Over existing
observatories, the optimum is that of Table~\ref{tab:layouts}. Over all candidates,
the largest gap stays at $33^{\circ}$, between New Zealand and Hawaii, because no
candidate at 900\,m or higher lies between them. The best layout found differs from
that of Fig.~\ref{fig:array14opt} by three sites, Teide, Skinakas and Ma\"{\i}do
instead of Pico, Pic du Midi and Tsiafajavona, and it needs only one site without
an observatory, Mt Marau on Tahiti. We repeat the additional searches with the
weight of 0.36, and they find the same layout. With this weight too, its $J$ is
slightly higher than that of the layout of Fig.~\ref{fig:array14opt}. The search
of Sect.~A.1 does not find it, because no single replacement of a site of
Fig.~\ref{fig:array14opt} raises $J$. Its coverage differs from that of
Fig.~\ref{fig:array14opt} by less than 0.2 percentage point, too little to change
any conclusion of this appendix.
With the minimum altitude of 1500\,m, the largest gap stays at $66^{\circ}$, between
Japan and Hawaii. The larger weight makes a difference only without a minimum
altitude: the
largest gap falls from $39^{\circ}$ to $31^{\circ}$, or 2.0\,h. The coverage, however,
falls by one percentage point, to 61.7\,\%, and five of the fourteen sites have no
observatory, three of them below 900\,m. With a minimum altitude, the largest gap is
therefore set by the geography of the Pacific, not by the weight of the penalty.

\subsection{Site list}

Table~\ref{tab:sites} lists the 60 candidate sites and the CTAO-South site.

\begin{table*}
\caption{The 60 candidate sites for the stations, ordered by longitude, and the CTAO-South site. Each site is named by its observatory or by a named point, in most cases a summit, followed by its island or region, where useful, and its country. Columns 2--4: latitude, longitude and altitude. Column~5 (Obs.): a tick marks an existing astronomical or astroparticle observatory, or the atmospheric research stations at Ma\"{\i}do and Pico Espejo; the other sites have no observatory. Column~6: role of the site in the layout of Fig.~\ref{fig:array} (core station, extension station or further station). Columns 7--10: a tick marks the sites chosen in each of the four optimised layouts of Table~\ref{tab:layouts}. These four layouts differ only in the sites that the optimisation was allowed to use: existing observatories at 900\,m or higher (Ex.); all candidates at 900\,m or higher (All; the layout of Fig.~\ref{fig:array14opt}); all candidates at any altitude (None); and all candidates at 1500\,m or higher (1500\,m).}
\label{tab:sites}
\centering
\footnotesize\setlength{\tabcolsep}{5pt}
\renewcommand{\arraystretch}{0.96}
\begin{tabular}{@{}l r r r c l c c c c@{}}
\toprule
Site & Lat. & Long. & Alt. & Obs. & Role in & \multicolumn{4}{c}{Optimised layouts}\\
\cmidrule(lr){7-10}
 & (deg) & (deg) & (m) & & Fig.~\ref{fig:array} & Ex. & All & None & 1500\,m\\
\midrule
Maungatere Hill, Chatham Islands, New Zealand & $-44.05$ & $-176.55$ & 294 &  &  &  &  & \checkmark & \\
Haleakal\={a}, Hawaii, USA & $+20.71$ & $-156.26$ & 3055 & \checkmark & core &  & \checkmark &  & \checkmark\\
Mauna Kea, Hawaii, USA & $+19.83$ & $-155.47$ & 4205 & \checkmark &  & \checkmark &  & \checkmark & \\
Mt Marau, Tahiti, France & $-17.62$ & $-149.53$ & 1493 &  &  &  & \checkmark &  & \\
Lick, California, USA & $+37.34$ & $-121.64$ & 1283 & \checkmark &  & \checkmark & \checkmark &  & \\
Palomar, California, USA & $+33.36$ & $-116.86$ & 1712 & \checkmark &  &  &  & \checkmark & \checkmark\\
San Pedro M\'artir, Mexico & $+31.03$ & $-115.49$ & 2830 & \checkmark &  &  &  &  & \\
Kitt Peak, Arizona, USA & $+31.96$ & $-111.60$ & 2096 & \checkmark &  &  &  &  & \\
Mt Hopkins, Arizona, USA & $+31.69$ & $-110.89$ & 2606 & \checkmark & core &  &  &  & \\
Maunga Terevaka, Easter Island, Chile & $-27.08$ & $-109.38$ & 507 &  &  &  &  & \checkmark & \\
Apache Point, New Mexico, USA & $+32.78$ & $-105.82$ & 2788 & \checkmark &  &  &  &  & \\
McDonald, Texas, USA & $+30.67$ & $-104.02$ & 2075 & \checkmark &  &  &  &  & \\
Sierra Negra, Mexico & $+18.99$ & $-97.31$ & 4600 & \checkmark &  & \checkmark & \checkmark &  & \checkmark\\
Cerro Crocker, Gal\'apagos, Ecuador & $-0.64$ & $-90.33$ & 864 &  &  &  &  &  & \\
Pico Espejo, Venezuela & $+8.53$ & $-71.05$ & 4765 & \checkmark &  &  &  &  & \\
Llano del Hato, Venezuela & $+8.79$ & $-70.87$ & 3600 & \checkmark & further &  &  & \checkmark & \checkmark\\
Cerro Tololo, Chile & $-30.17$ & $-70.81$ & 2200 & \checkmark &  &  &  &  & \checkmark\\
Cerro Pach\'on, Chile & $-30.24$ & $-70.74$ & 2715 & \checkmark &  &  &  &  & \\
Paranal, Chile & $-24.63$ & $-70.40$ & 2635 & \checkmark &  & \checkmark & \checkmark &  & \\
CTAO-South, Chile & $-24.68$ & $-70.32$ & 2150 & \checkmark &  &  &  &  & \\
El Leoncito, Argentina & $-31.80$ & $-69.30$ & 2552 & \checkmark & core &  &  &  & \\
Pico dos Dias, Brazil & $-22.53$ & $-45.58$ & 1864 & \checkmark & ext & \checkmark & \checkmark & \checkmark & \checkmark\\
Pico, Azores, Portugal & $+38.47$ & $-28.40$ & 2351 &  &  &  & \checkmark &  & \checkmark\\
Fogo, Cape Verde & $+14.95$ & $-24.34$ & 2829 &  &  &  &  &  & \\
ORM, La Palma, Spain (CTAO-North) & $+28.76$ & $-17.89$ & 2200 & \checkmark & core &  &  & \checkmark & \\
Teide, Tenerife, Spain & $+28.30$ & $-16.51$ & 2390 & \checkmark &  & \checkmark &  &  & \\
Green Mountain, Ascension, UK & $-7.95$ & $-14.35$ & 859 &  &  &  &  &  & \\
Diana's Peak, St Helena, UK & $-15.96$ & $-5.69$ & 818 &  &  &  &  &  & \\
Calar Alto, Spain & $+37.22$ & $-2.55$ & 2168 & \checkmark &  &  &  &  & \\
Pic du Midi, France & $+42.94$ & $+0.14$ & 2877 & \checkmark &  &  & \checkmark &  & \checkmark\\
H.E.S.S. site, Namibia & $-23.27$ & $+16.50$ & 1800 & \checkmark & core & \checkmark & \checkmark & \checkmark & \checkmark\\
Sutherland, South Africa & $-32.38$ & $+20.81$ & 1798 & \checkmark &  &  &  &  & \\
Skinakas, Crete, Greece & $+35.21$ & $+24.90$ & 1750 & \checkmark &  & \checkmark &  & \checkmark & \\
Kottamia, Egypt & $+29.93$ & $+31.83$ & 476 & \checkmark &  &  &  &  & \\
Abastumani, Georgia & $+41.75$ & $+42.82$ & 1650 & \checkmark &  &  &  &  & \checkmark\\
Mt Aragats, Armenia & $+40.47$ & $+44.18$ & 3200 & \checkmark & core &  &  &  & \\
Tsiafajavona, Madagascar & $-19.37$ & $+47.24$ & 2643 &  &  &  & \checkmark &  & \\
Ma\"{\i}do, R\'eunion, France & $-21.08$ & $+55.38$ & 2160 & \checkmark & further & \checkmark &  & \checkmark & \checkmark\\
Maidanak, Uzbekistan & $+38.67$ & $+66.90$ & 2593 & \checkmark &  &  &  &  & \\
Port-aux-Fran\c{c}ais, Kerguelen, France & $-49.35$ & $+70.22$ & 10 &  &  &  &  &  & \\
Mt Abu, India & $+24.65$ & $+72.78$ & 1680 & \checkmark &  &  &  &  & \checkmark\\
Martin-de-Vivi\`es, Amsterdam Island, France & $-37.78$ & $+77.57$ & 29 &  &  &  &  &  & \\
Assy-Turgen, Kazakhstan & $+43.23$ & $+77.87$ & 2750 & \checkmark &  &  &  &  & \\
Hanle, India & $+32.78$ & $+78.96$ & 4500 & \checkmark &  & \checkmark & \checkmark &  & \\
Devasthal, India & $+29.36$ & $+79.68$ & 2450 & \checkmark &  &  &  & \checkmark & \\
Ali, Tibet, China & $+32.32$ & $+80.02$ & 5100 & \checkmark & core &  &  &  & \\
Yangbajing, Tibet, China & $+30.11$ & $+90.53$ & 4300 & \checkmark &  &  &  &  & \\
Gaomeigu, Yunnan, China & $+26.70$ & $+100.03$ & 3200 & \checkmark & ext &  &  &  & \checkmark\\
Bosscha, Indonesia & $-6.82$ & $+107.62$ & 1310 & \checkmark &  & \checkmark & \checkmark &  & \\
Murchison, Australia & $-26.70$ & $+116.63$ & 361 & \checkmark &  &  &  & \checkmark & \\
Xinglong, China & $+40.39$ & $+117.58$ & 960 & \checkmark &  &  &  &  & \\
Timau, Indonesia & $-9.60$ & $+123.95$ & 1347 & \checkmark &  &  &  &  & \\
Bohyunsan, South Korea & $+36.16$ & $+128.98$ & 1143 & \checkmark &  &  &  & \checkmark & \\
Norikura, Japan & $+36.11$ & $+137.55$ & 2770 & \checkmark & core & \checkmark & \checkmark &  & \checkmark\\
Mt Lamlam, Guam, USA & $+13.34$ & $+144.66$ & 406 &  &  &  &  &  & \\
Mt Stromlo, Australia & $-35.32$ & $+149.01$ & 770 & \checkmark &  &  &  &  & \\
Siding Spring, Australia & $-31.27$ & $+149.06$ & 1165 & \checkmark & core & \checkmark &  &  & \\
Mont Do, New Caledonia, France & $-21.75$ & $+166.00$ & 1024 &  &  &  &  &  & \\
Wake Island, USA & $+19.30$ & $+166.63$ & 6 &  &  &  &  & \checkmark & \\
Mt John, New Zealand & $-43.99$ & $+170.46$ & 1029 & \checkmark & core & \checkmark & \checkmark &  & \\
Nadi, Fiji & $-17.76$ & $+177.44$ & 18 &  &  &  &  &  & \\
\bottomrule
\end{tabular}
\end{table*}

\section{Results for other zenith-angle limits}
\label{app:zenith}

\begin{figure*}[t]
\centering
\includegraphics[width=\textwidth]{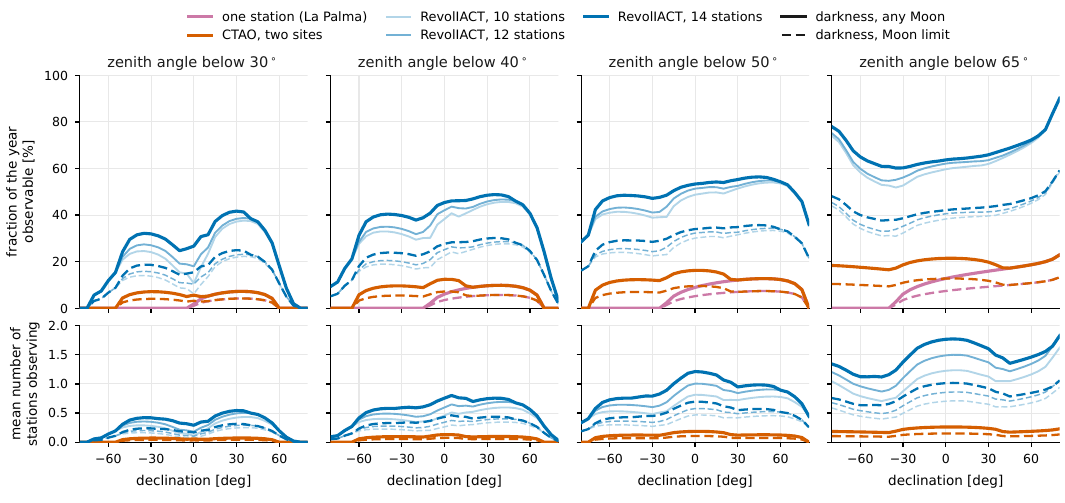}
\caption{Annual temporal coverage as a function of source declination, as in
Fig.~\ref{fig:coverage}, for zenith-angle limits of $30^{\circ}$, $40^{\circ}$,
$50^{\circ}$ and $65^{\circ}$ (from left to right; $60^{\circ}$ is shown in
Fig.~\ref{fig:coverage}). Top: fraction of the year during
which at least one station can observe the source. Bottom: mean number of stations
observing it simultaneously. Solid curves (`darkness, any Moon') require
astronomical darkness; dashed curves (`darkness, Moon limit') require in addition a
Moon below the horizon or less than 40\,\% illuminated.}
\label{fig:coverage_zenith}
\end{figure*}

In the main text we use a zenith-angle limit of $60^{\circ}$. A larger limit gives more
observing time, because each source stays within it for more hours per night. The
energy threshold of a station, however, rises with the zenith angle
(Sect.~\ref{sec:sensitivity}), so that observations at large zenith angles miss the
lowest energies. The best limit therefore depends on the source. A strict limit
suits distant sources and sources with soft spectra, which emit most of their gamma
rays at the lowest energies. A larger limit suits bright sources with hard spectra,
for which the observing time matters more than the lowest energies. Here we repeat
the main calculations of Sects.~\ref{sec:coverage} and \ref{sec:grb} for limits of $30^{\circ}$,
$40^{\circ}$, $50^{\circ}$ and $65^{\circ}$, with all other conditions unchanged.
Figure~\ref{fig:coverage_zenith} shows the coverage as a function of declination,
as Fig.~\ref{fig:coverage} does for $60^{\circ}$, and Table~\ref{tab:zenith}
compares the main results.

\begin{table*}[t]
\caption{Main results for zenith-angle limits from $30^{\circ}$ to $65^{\circ}$; the column for $60^{\circ}$ repeats those of the main text. Energy threshold: factor by which it exceeds its value near the zenith, from the scaling $(\cos\theta_{\rm z})^{-2.3}$ found for MAGIC \citep{aleksic2016}, where $\theta_{\rm z}$ is the zenith angle, and the resulting threshold of a station with a threshold of 0.075\,TeV near the zenith. Coverage: fraction of the year during which at least one station can observe a source, averaged over the sky as in Sect.~\ref{sec:coverage}, in darkness and, in parentheses, with the Moon limit. Source at opposition: the twenty cases of Sect.~\ref{sec:coverage}, in darkness. GBM bursts: fraction of the bursts of Table~\ref{tab:grb} observable at the trigger, in darkness.}
\label{tab:zenith}
\centering
\small\setlength{\tabcolsep}{6pt}
\begin{tabular}{l c c c c c}
\toprule
Zenith-angle limit & $30^{\circ}$ & $40^{\circ}$ & $50^{\circ}$ & $60^{\circ}$ & $65^{\circ}$\\
\midrule
\multicolumn{6}{l}{\textit{Energy threshold}}\\
\quad Relative to the zenith & 1.4 & 1.8 & 2.8 & 4.9 & 7.2\\
\quad For 0.075\,TeV near the zenith (TeV) & 0.10 & 0.14 & 0.21 & 0.37 & 0.54\\
\multicolumn{6}{l}{\textit{Sky-averaged coverage (\%), darkness (Moon limit)}}\\
\quad La Palma alone & 2.6 (1.5) & 4.5 (2.6) & 6.8 (3.9) & 9.5 (5.5) & 10.9 (6.3)\\
\quad CTAO (North and South) & 5.1 (2.9) & 8.9 (5.1) & 12.9 (7.5) & 17.3 (10.1) & 19.4 (11.4)\\
\quad 10 core stations & 22.5 (13.1) & 35.4 (20.9) & 46.2 (27.9) & 56.1 (34.7) & 60.8 (38.1)\\
\quad 12 stations (10+2) & 24.7 (14.5) & 37.4 (22.4) & 48.1 (29.6) & 58.0 (36.6) & 62.7 (40.0)\\
\quad 14 stations (10+2+2) & 28.3 (16.6) & 40.7 (24.6) & 51.2 (31.9) & 60.8 (38.9) & 65.4 (42.4)\\
\quad Gain of 14 stations over CTAO & 5.5 (5.6) & 4.6 (4.8) & 4.0 (4.2) & 3.5 (3.9) & 3.4 (3.7)\\
\multicolumn{6}{l}{\textit{Source on the celestial equator, darkness}}\\
\quad Hours per day, CTAO & 1.4 & 3.0 & 3.9 & 4.7 & 5.2\\
\quad Hours per day, 14 stations & 6.4 & 10.9 & 12.8 & 14.5 & 15.3\\
\quad Mean number of the 14 stations observing & 0.29 & 0.76 & 1.21 & 1.59 & 1.77\\
\quad Station-hours per day, 14 stations & 7 & 18 & 29 & 38 & 42\\
\quad Sensitivity in one day, 14 stations (\% C.U.)\tablefootmark{c} & 5.7 & 3.5 & 2.8 & 2.4 & 2.3\\
\quad Sensitivity in one day, IACT system on La Palma (\% C.U.)\tablefootmark{c} & 7.5 & 4.1 & 3.4 & 3.0 & 2.9\\
\multicolumn{6}{l}{\textit{Source at opposition (20 cases)\tablefootmark{a}}}\\
\quad Cases covered around the clock, 14 stations & 4 & 8 & 18 & 18 & 19\\
\quad Shortest daily coverage, 14 stations (h) & 6.2 & 16.6 & 21.0 & 21.0 & 21.0\\
\quad Daily coverage, CTAO (h)\tablefootmark{b} & 0.0--4.6 & 3.6--7.6 & 6.9--9.3 & 6.9--10.9 & 6.9--11.7\\
\multicolumn{6}{l}{\textit{GBM bursts observable at the trigger (\%), darkness}}\\
\quad CTAO (North and South) & 5.1 & 8.9 & 12.5 & 16.6 & 18.9\\
\quad 14 stations (10+2+2) & 28.4 & 40.6 & 50.9 & 61.2 & 66.4\\
\quad Gain of 14 stations over CTAO & 5.6 & 4.6 & 4.1 & 3.7 & 3.5\\
\bottomrule
\end{tabular}
\tablefoot{
\tablefoottext{a}{A source at opposition, that is, opposite the Sun, on four dates, the equinoxes and solstices of 2027, and at five declinations, $-60^{\circ}$, $-30^{\circ}$, $0^{\circ}$, $+30^{\circ}$ and $+60^{\circ}$ (Sect.~\ref{sec:coverage}).}
\tablefoottext{b}{The value 0.0 means that, with a limit of $30^{\circ}$, neither CTAO site can observe the sources at $-60^{\circ}$ and $+60^{\circ}$: from La Palma ($28.8^{\circ}$\,N) and from the CTAO-South site ($24.7^{\circ}$\,S), they never come within $30^{\circ}$ of the zenith.}
\tablefoottext{c}{From Eq.~(\ref{eq:tdet}), for a source on the celestial equator, with $\sens{15}$ per station (data of all stations combined) and $\sens{5}$ for a current IACT system (Sect.~\ref{sec:sensitivity}); the rise of the energy threshold with the zenith angle is neglected.}}
\end{table*}

\paragraph{Energy threshold.} With the scaling of Sect.~\ref{sec:sensitivity},
$(\cos\theta_{\rm z})^{-2.3}$, the threshold is 1.4, 1.8, 2.8, 4.9 and 7.2 times
higher at $30^{\circ}$, $40^{\circ}$, $50^{\circ}$, $60^{\circ}$ and $65^{\circ}$
than near the zenith. A station with a threshold of 0.075\,TeV near the zenith
therefore reaches 0.10, 0.14, 0.21, 0.37 and
0.54\,TeV. Up to $50^{\circ}$, its threshold stays near or below 0.2\,TeV, the
energy above which Table~\ref{tab:cases} quotes the fluxes.

\paragraph{Coverage.} A stricter limit reduces the coverage of every configuration,
but not by the same amount (Fig.~\ref{fig:coverage_zenith}). Averaged over the sky,
the fourteen stations observe a source 28\,\% of the time with a limit of
$30^{\circ}$, 41\,\% with $40^{\circ}$, 51\,\% with $50^{\circ}$, 61\,\% with
$60^{\circ}$ and 65\,\% with $65^{\circ}$ (Table~\ref{tab:zenith}). The two CTAO sites lose relatively more:
their coverage falls from 17\,\% at $60^{\circ}$ to 13\,\% at $50^{\circ}$ and
5\,\% at $30^{\circ}$. With a limit of $30^{\circ}$, a single site sees only a band
of declinations $60^{\circ}$ wide, centred on its latitude. The stations of the
array, between latitudes $44^{\circ}$\,S and $40^{\circ}$\,N, together still cover
the declinations from about $-74^{\circ}$ to $+70^{\circ}$. The gain of the fourteen
stations over CTAO therefore grows as the limit becomes stricter: from 3.4 at
$65^{\circ}$ and 3.5 at $60^{\circ}$ to 4.0 at $50^{\circ}$ and 5.5 at $30^{\circ}$,
and slightly more with the Moon limit. These gains assume the same limit for CTAO.
CTAO, however, suffers much less than a compact station from observations at large
zenith angles. The large telescopes of the CTAO LST sub-array keep an energy
threshold below 0.1\,TeV up to zenith angles of $40$--$50^{\circ}$
(Sect.~\ref{sec:zenith65}), so that CTAO loses little of the lowest energies when it
observes up to these angles. Therefore, with strict zenith-angle limits, such as
$30^{\circ}$, the gains reported here overstate what the array of compact telescopes
adds to CTAO.

\paragraph{Sources near the celestial poles.} With limits of $60^{\circ}$ and
$65^{\circ}$, the coverage rises towards both celestial poles, whereas with
stricter limits it falls (Figs.~\ref{fig:coverage} and \ref{fig:coverage_zenith}).
For a single station, the coverage averaged over right ascension is the fraction of
the year during which the station is dark, times the fraction of the day during
which the source is within the zenith-angle limit. A source near the celestial
equator rises and sets. It is within the limit only for a few hours around its
transit: with a limit of $65^{\circ}$, for 34\,\% of the day at a latitude of
$30^{\circ}$. A source near a celestial pole instead circles the pole, and its
zenith angle changes little during the day. It therefore stays within a large
limit for much of the day, or for all of it. For La Palma alone, with a limit of
$60^{\circ}$, the coverage rises from 11.6\,\% on the celestial equator to
16.8\,\% at a declination of $+80^{\circ}$. With a limit of $65^{\circ}$, a source
at $+80^{\circ}$ never leaves the limit as seen from Mt Aragats and Norikura, and
a source at $-80^{\circ}$ never leaves it as seen from Mt John. These sources can
be observed whenever one of these stations is dark, and the fourteen stations
cover them for 90\,\% and 78\,\% of the year, respectively. With a limit of
$50^{\circ}$ or less, most stations cannot see these sources at all, and the
coverage falls towards the poles.

\paragraph{Simultaneous observations.} The mean number of stations that observe a
source at the same time falls faster than the coverage. For a source on the
celestial equator, it is 1.6 at $60^{\circ}$, 1.2 at $50^{\circ}$, 0.8 at
$40^{\circ}$ and 0.3 at $30^{\circ}$, or 38, 29, 18 and 7 station-hours per day.
The sensitivity of the array in one day, with the data of the stations combined
(Sect.~\ref{sec:sensitivity}), is therefore about 2.4\,\%~C.U.\ at $60^{\circ}$ and
2.8\,\%~C.U.\ at $50^{\circ}$, compared with 3.0 and 3.4\,\%~C.U.\ for a current IACT
system on La Palma (Table~\ref{tab:zenith}).

\paragraph{Continuity of observation for a source at opposition.} The continuous
observation described in Sect.~\ref{sec:coverage} needs a limit of about $50^{\circ}$
or more. With fourteen stations, at least one station can observe a source at
opposition around the clock in 18 of the 20 cases of Sect.~\ref{sec:coverage}
with limits of $50^{\circ}$ and $60^{\circ}$, and in 19 with $65^{\circ}$. With
$40^{\circ}$, this is true in 8 cases, and in every case the source can be observed
for at least 16.6 hours per day. With $30^{\circ}$, it is true in 4 cases, and in
the worst case the source can be observed for only 6.2 hours per day. The two CTAO
sites can observe a source at opposition for at most 4.6 hours per day with a limit
of $30^{\circ}$, and for at most 11.7 hours with $65^{\circ}$.

\paragraph{GBM bursts.} The fraction of the GBM bursts observable at the trigger
depends on the limit in the same way as the sky-averaged coverage
(Table~\ref{tab:zenith}). It falls for every configuration as the zenith-angle limit
becomes stricter, but relatively less for the array than for CTAO. The gain of the fourteen
stations over CTAO therefore grows, from \GbmGainDarkSixtyFive{} at $65^{\circ}$ to
\GbmGainThirty{} at $30^{\circ}$, and it stays within 5\,\% of the gain in
sky-averaged coverage at every limit.

\paragraph{Which limit to choose.} The choice depends on the science. Below
$30^{\circ}$, the threshold stays close to its value at the zenith, which favours
distant sources and soft spectra (Sect.~\ref{sec:sensitivity}), but the coverage is
less than half of that at $60^{\circ}$. Up to $50^{\circ}$, a station with a threshold of 0.075\,TeV at the zenith keeps its
threshold near 0.2\,TeV. The coverage of the
fourteen stations is then 51\,\%, and a source at opposition can still be followed
around the clock in 18 of the 20 cases. Beyond $50^{\circ}$, the threshold rises
above 0.2\,TeV, and the additional coverage serves mainly bright sources with hard
spectra. The limit can be chosen source by source. At every limit, the array offers
3.4 to 5.5 times the coverage of the two CTAO sites with the same limit.

\end{appendix}

\end{document}